\documentclass[useAMS,usenatbib]{mn2e}
\usepackage{xspace}
\usepackage[]{natbib}
\usepackage{times}
\usepackage{amsmath} 
\usepackage{amssymb} 
\usepackage{graphicx} 
\usepackage{indentfirst} 
\usepackage{pdflscape} 
\usepackage{placeins} 
\usepackage{nicefrac} 
\usepackage{afterpage} 
\usepackage{rotating} 
\usepackage{booktabs} 
\usepackage{rotating} 
\usepackage{color} 
\usepackage{caption}

\newcommand{\apl}{\:^{<}_{\sim}\:}

\newcommand{\kms}{\mbox{\,km\ s${^{-1}}$}}

 \newcommand{\nab}{\mbox{\boldmath$\nabla$}}
 \newcommand{\vel}{\mbox{\boldmath$v$}}

\begin{document}

\title[Column Density Profiles of Gaseous Halos]{Column Density Profiles of Multi-Phase Gaseous Halos}

\author[Liang et al. ]{Cameron J. Liang$^{1}$\thanks{E-mail:jwliang@oddjob.uchicago.edu}, Andrey V. Kravtsov$^{1,2}$, Oscar Agertz$^{3}$\\
\\
$^{1}$Department of Astronomy \& Astrophysics, and Kavli Institute for Cosmological Physics, University of Chicago, Chicago IL 60637 \\
$^{2}$Enrico Fermi Institute, The University of Chicago, Chicago, IL 60637, USA \\
$^{3}$Department of Physics, University of Surrey, Guildford, GU2 7XH, UK}

\pagerange{\pageref{firstpage}--\pageref{lastpage}} \pubyear{2015}

\maketitle

\label{firstpage}

\begin{abstract}
We analyze circumgalactic medium (CGM) in a suite of high-resolution cosmological re-simulations of a Milky-Way size galaxy and show that CGM properties are quite sensitive to details of star formation--feedback loop modelling. The simulation that produces a realistic late-type galaxy, fails to reproduce existing observations of the CGM. In contrast, simulation that does not produce a realistic galaxy has the predicted CGM in better agreement with observations. This illustrates that properties of galaxies and properties of their CGM provide strong {\it complementary} constraints on the processes governing galaxy formation. Our simulations predict that column density profiles of ions are well described by an exponential function of projected distance $d$: $N \propto e^{-d/h_s}$. Simulations thus indicate that the sharp drop in absorber detections at larger distances in observations does not correspond to a ``boundary'' of an ion, but reflects the underlying steep exponential column density profile. 
Furthermore, we find that ionization energy of ions is tightly correlated with the scale height $h_s$:  $h_s \propto E_{\rm ion}^{0.74}$.  At $z \approx 0$, warm gas traced by low-ionization species (e.g., Mg\,II and C\,IV) has $ h_s \approx 0.03-0.07 R_{\rm vir}$, while higher ionization species (O\,VI and Ne\,VIII) have $h_s \approx 0.32-0.45R_{\rm vir}$.  Finally, the scale heights of ions in our simulations evolve slower than the virial radius for $z\leq 2$, but similarly to the halo scale radius, $r_s$. Thus, we suggest that the column density profiles of galaxies at different redshifts should be scaled by $r_s$ rather than the halo virial radius.  
\end{abstract}

\begin{keywords}
cosmology:theory -- galaxies:halos -- simulations:feedback
\end{keywords}

%------------------------
\section{Introduction}
%------------------------

A complete picture of galaxy evolution requires an understanding of the interplay between inflow and cooling of gas, star formation, and associated feedback that can drive star-forming gas
back into the halo or beyond via large-scale winds. The current generation of galaxy formation models appears to capture many of these processes in a way that leads to galaxies with realistic properties \citep[e.g.,][]{Feldmann2011,Sales2012,Stinson2013,Aumer2013,Hopkins2014,AgertzKravtsov2015,Schaye2015}. So far, however, galaxy formation models have been tested almost exclusively against properties of the stellar component of galaxies \citep[with some recent exceptions, e.g.,][]{Hummels2013,Ford2013,Ford2015,Schaye2015}. 

Yet, we know that stars comprise  only a relatively small fraction of the baryon budget. A variety of observational constraints and inferences from the abundance matching technique \citep[e.g., see Fig. 10 and 11 in][for a compilation of recent constraints]{Kravtsov2014} shows that even the galaxies that are most efficient in forming stars (forming in halos of $M_h \approx 10^{12} M_{\odot}$)  convert at most $\approx 30-40\%$ of the available baryon budget into stars. The bulk of the remainder of the baryon budget is outside of the disk, because cold gas fractions in the evolved galaxies we observe today are also small.   The ``missing'' baryons are thought to reside in the the circumgalactic medium (CGM) around galaxies, which can be thought of as a halo of tenuous gas resulting from the complex interactions between static gaseous halo, inflowing gas, and feedback-driven outflows from galaxies. Studying  the galactic baryons in the CGM is therefore paramount for the understanding of the feedback processes that regulate inflows and outflows of gas. 

Although details of such processes are still debated, efficient feedback recipes have been implemented in galaxy formation simulations \citep[e.g.,][]{Hopkins2011, Brook2012,Stinson2013,Vogelsberger2013,Agertz2013, Roskar2014,TrujilloGomez2015}.    At the low mass end ($M_h < 10^{12} M_{\odot}$), stellar feedback in the form of energy and momentum injection from radiation pressure, stellar winds, and supernova explosions  is believed to be responsible for the suppression of star formation \citep{DekelSilk1986,Efstathiou2000}. On the other hand, active galactic nuclei (AGN) feedback is thought to be critical in limiting star formation and stellar masses of galaxies at the high mass end \citep[$M_h > 10^{12} M_{\odot}$; e.g., ][]{SilkRees1998, Benson2003}. More recently, cosmic ray (CR) driven winds have also been shown to be a promising feedback mechanism in the context of galaxy formation simulations \citep[e.g.,][]{Socrates2008,Booth2013, Salem2014}. 

It is quite likely that stellar feedback leaves a specific imprint on the properties of the circumgalactic gaseous halos \citep[e.g.,][]{Barai2013, Hummels2013,Suresh2015}. In particular, comparisons between simulated CGM and observations have indicated that the observed incidence of absorbers in the CGM favors larger intensity of feedback processes \citep[e.g.,][]{Stinson2012, Hummels2013,Suresh2015}. 

Observations of the CGM have advanced tremendously in recent years. The Cosmic Origins Spectrograph \citep[COS;][]{Green2012} on board the \textit{Hubble Space Telescope (HST)} has offered a unique opportunity to systematically probe low-redshift ($z \approx 0 $) galactic halos in the UV regime. Commonly observed absorption lines arising in the CGM are the Ly$\alpha$ $\lambda$1215, Si\,II $\lambda$1260, Si\,III $\lambda$1206, C\,II $\lambda$1334, C\,IV $\lambda$$\lambda$1548, 1550 \citep[e.g.,][]{Chen2012, Borthakur2013, LiangChen2014, Bordoloi2014}, and O\,VI $\lambda$$\lambda$1031, 1036 \citep[e.g.,][]{Prochaska2011, Tumlinson2011, Johnson2015}. In the intermediate redshift range, many researchers have probed the gaseous halos using optical ground-based facilities via Mg\,II \citep{Chen2010,Gauthier2010,Bordoloi2011}. Similar studies for the UV transitions have also been conducted in the high redshift universe \citep[$z \approx 2$;][]{Steidel2010, Rudie2012,Turner2014}. 

In this paper, we explore how the choices of parameters controlling star formation and stellar feedback in cosmological galaxy formation simulations affect the properties of the CGM around simulated galaxies, such as the profile of the absorber column densities. We also use the simulation in which the CGM properties are closest to  observations to explore the physical origin of the CGM absorbers and the processes that shape their properties. 

The paper is structured as follows. In section 2, we outline the simulation suite used in our study, including details of star formation and the stellar and cosmic ray feedback models. In section 3, we describe the post-processing of simulations and our CGM analyses. Main qualitative features of the CGM in our simulations are discussed in section 4, where we demonstrate that the radial profiles of the CGM absorber column density have an exponential form and evolve only very weakly with redshift. In addition, we show that profiles of the CGM spanning four decades of stellar mass and 11 billion years in cosmic time can be cast as a single profile using a simple re-scaling of radius ($d\rightarrow d/r_s$). We then present comparisons between our predictions from simulations and observations of the CGM in section 5, where we show that observations also appear to obey scaling with halo scale radius. In section 6 we discuss our results and their implications, as well as compare them to results of other recent studies of the CGM in simulations. Finally, in section 7, we summarize our results and conclusions. 

%---------------------------------------------
\section{Galaxy Formation Simulations}
%--------------------------------------------

In this study, we use a suite of galaxy formation simulations, which consists of ``zoom-in'' re-simulations of the evolution of a Milky Way-sized halo started from the same initial conditions, but with different parameters of star formation and stellar feedback recipes, as described in \cite{AgertzKravtsov2015}. In addition, we use a new re-simulation with a cosmic ray feedback model described in \cite{Booth2013}. 

All simulations are run using the Adaptive Mesh Refinement (AMR) code {\tt RAMSES} \citep{Teyssier2002} from the same initial conditions in the WMAP5 $\rm \Lambda CDM$ cosmology with $\Omega_{\Lambda} = 0.73$, $\Omega_{\rm m} = 0.27$, $\Omega_{\rm b} = 0.045$, $\sigma_8 = 0.8$ and $H_0 = 70 \kms \rm Mpc^{-1}$ \citep{Komatsu2009}. At $z=0$, the halo masses of the simulated $\sim L_*$ galaxies have similar values of $M_{\rm vir} \approx 10^{12} M_{\odot}$, corresponding to a virial radius of $R_{\rm vir} \approx 260$ kpc. The dark matter particle mass in the highest resolution region is $m_{\rm DM} = 3.2 \times 10^5 M_{\odot}$. The physical resolution at maximum refinement level reaches $\Delta x \approx 75$ pc.  A summary of the parameters of our simulations is provided in Table 1. 

Unless otherwise stated, all quantities in the paper are given in physical units.   We briefly outline the main ingredients of the simulations in the following subsections. A more detailed description can be found in \cite{AgertzKravtsov2015}, while the specifics on the implementation of stellar and cosmic ray feedback are detailed in \cite{Agertz2013} and \cite{Booth2013}, respectively. 

%%%%%%%%%%%%%%%%%%%%%%%%%%%%%%%%%%%%%%%%%%%%%%%%
\begin{footnotesize}
\begin{table*}
\begin{center}
 \footnotesize
\centering
%    \hspace*{-1.85 cm}
\begin{minipage}{160mm}
   \caption{Simulation suite}
   \begin{tabular}{@{}lllc@{}}
     \hline
     \hline
	Simulations 			&  Description	 & Final Redshift \\
     \hline
     KMT09 models, feedback energy variable $E_{\rm{fb}}$, $f_{\rm{fb}} = 0.5$,  $t_{\rm{dis}} = 10$Myr  \\
     \hline
     \texttt{ALL\_Efb\_e010} 			&  All feedback processes, $\epsilon_{\rm{ff}} = 10\% $	 & $z = 0$ \\
     \texttt{ALL\_Efb\_e001} 			&  All feedback processes, $\epsilon_{\rm{ff}} = 1\% $	 & $z = 1.5$ \\
     \texttt{ALL\_Efb\_e001\_5ESN} 			&  All feedback processes,  $E_{\rm{SNII}} = 5\times 10^{51}$ erg, $\epsilon_{\rm{ff}} = 1\%$ & $z = 0$ \\
     \hline
     Cosmic rays model, feedback energy variable $E_{\rm CR} = \xi_{\rm{CR}} E_{\rm SNII}$   \\
     \hline
     \texttt{ALL\_e010\_CR} 			&  All feedback processes, $\xi_{\rm{CR}} = 10 \%$, $\epsilon_{\rm{ff}} = 10\% $ & $z = 1.0$ \\
     \hline    

   \label{line_table}    
 \end{tabular}
\end{minipage}
\end{center}
\end{table*}
\end{footnotesize}
%%%%%%%%%%%%%%%%%%%%%%%%%%%%%%%%%%%%%%%%%%%%%%%%

\subsection{Star Formation} 

In each star forming cell, the number of star particles $N$ is determined from the Poisson distribution $P(N | \lambda_P)$ with the mean $\lambda_P = \dot{\rho}_* \Delta x^3\Delta t/m_*$, where star formation rate is given by:
\begin{equation} \dot{\rho}_* = f_{\rm{H_2}} \frac{\rho_{\rm{g}}}{t_{\rm{ff}}} \epsilon_{\rm{ff}}.  \end{equation} 
Here $f_{\rm{H_2}}$ is the local mass fraction of molecular hydrogen computed using the model of \cite{Krumholz2009}, as described in \cite{AgertzKravtsov2015}, $\rho_{\rm{g}}$ is the local gas density of a given cell, $t_{\rm{ff}}=\sqrt{3\pi/(32G\rho_{\rm g})}$ is the free-fall time, and $\epsilon_{\rm{ff}}$ is the star formation efficiency per free-fall time assumed to be constant in time and space. The simulations analyzed in this paper use two constant values of the efficiency: $\epsilon_{\rm{ff}} = 1\%$ and $10\%$.  
The parameter $m_*$, set to $10^4 M_{\odot}$ in our simulations, is a unit mass of star particles and all star particles are created as a multiple of it \citep{RaseraTeyssier2006, DuboisTeyssier2008}.

Parameters of the star formation recipe have a direct impact on the CGM because the feedback strength depends on the mass and spatial distribution of young stars, as we will discuss in section \ref{sec:cgmsim} \citep[see also][]{AgertzKravtsov2015, Stinson2012}.
  
\subsection{Stellar Feedback} 

Our simulations account for energy, momentum, mass, and the injection of heavy elements (with an effective yield of 1\% - 3\%) from type Ia supernovae (SNIa), type II supernovae (SNII), stellar winds and radiation pressure from massive stars, as well as for the secular mass loss into the surrounding interstellar medium (ISM) from the AGB stars.  Injection of momentum, energy and mass is done continuously during each simulation time step, which is computed according to the Courant-Friedrichs-Lewy (CFL) condition. In the regions of active star formation and feedback the time steps become as short as a few thousand years.

When a supernova explodes, we retain a fraction, $f_{\rm fb}$, of the explosion energy in a separate energy variable $E_{\rm fb}$, which is evolved using an equation similar to that of the internal energy of the gas, but with the explicitly specified dissipation time scale,  $t_{\rm dis}$:

\begin{equation}
\frac{\partial}{\partial t}(E_{\rm fb})+\nab\cdot(E_{\rm fb}\vel_{\rm gas} )=-P_{\rm fb}\nab\cdot\vel_{\rm gas}-\frac{E_{\rm fb}}{t_{\rm dis}} \label{eqn:EfbEqn}
\end{equation}

This energy variable is used to define an additional contribution to gas pressure, $P_{\rm fb} = (\gamma - 1) E_{\rm fb}$, which can be considered 
as a crude approximation  for effective pressure of the hot gas \citep{AgertzKravtsov2015} or subgrid turbulence unresolved in simulations and can allow for more efficient transfer of supernova momentum and energy into the surrounding ISM. The remaining $1 - f_{\rm fb}$ fraction of energy is added to the thermal energy of the gas in the cell containing the stellar particles. 
The dissipation time scale is adopted to be $t_{\rm dis}=10$ Myr for all simulations, comparable to a couple crossing times in massive giant molecular clouds. 

In addition, we model momentum injection due to radiation pressure, stellar winds, and supernovae by directly adding momentum into surrounding cells using the rate: 

\begin{equation} \dot{p}_{\rm{rad}} = (\eta_1 + \eta_2 \tau_{\rm IR}) \frac{L(t)}{c}, \end{equation} 

where $\tau_{\rm{IR}}$ is the optical depth in the infrared (IR) band, $L(t)$ is the luminosity of the stellar population, taken from the stellar evolution
code {\tt STARBURST99} \citep{Leitherer1999}. The first term corresponds to direct momentum injection from UV photons, where $\eta_1 $ is chosen to be unity. The second term corresponds to momentum transfer by IR photons re-emitted by dust after the UV photons are absorbed, where $\eta_2 = 2$ is adopted in \cite{AgertzKravtsov2015}. 

\subsection{Cosmic Ray Feedback} 

Simulations of isolated disk galaxies in \cite{Booth2013} and \cite{Salem2014} show that cosmic ray (CR) driven winds contain more cool/warm gas ( $T < 10^5 $ K)  than the winds driven via direct momentum and thermal energy injection from supernovae. This is because in the CR-driven winds  the gas is gradually accelerated by a large-scale pressure gradient established by the cosmic rays diffusing out of the disk, while in the standard winds produced by energy and momentum injection at SN sites the gas  is launched at large velocity and is then shock-heated to high temperatures. 
Therefore, we also consider a new re-simulation of the same halo as in the other simulations, but instead incorporates the CR feedback using the isotropic diffusion model described in \cite{Booth2013}.  

Briefly,  the CRs are assumed to be produced by supernovae with a fraction $\xi_{\rm CR}$ of the SN energy converted into the energy of cosmic rays: $E_{\rm CR} = \xi_{\rm{CR}}E_{\rm{SN}}$.  The rest of  $E_{\rm SN}$, is added to the thermal energy of the gas in the SN parent cell.  In the simulation used in this analysis, we adopted $\xi_{\rm{CR}} = 10\%$ consistent both with recent empirical evidence for CR acceleration in the SN remnants \citep{Ackermann2013} and theoretical models \citep{Malkov2001,CaprioliSpitkovsky2014}. 

The cosmic rays are modelled as an ultra-relativistic ideal fluid with $\gamma_{\rm{CR}} = 4/3$, which exerts pressure $P_{\rm{CR}} = (\gamma_{\rm{CR}} - 1) E_{\rm{CR}}$, where $E_{\rm{CR}}$ is the energy density of the cosmic ray particles followed using a separate variable. The variable is evolved using the advection-diffusion equation, which includes terms corresponding to cooling losses due to decays and Coulomb interactions of CR with gas and magnetic fields, an isotropic diffusion term assuming the constant diffusion coefficient of $\kappa=3\times 10^{-27}\ \rm cm^2\, s^{-1}$, and a source term due to the SN remnant production of cosmic rays. We also include the heating term into the energy equation of baryon gas due to interactions with CRs \citep[see details in][]{Booth2013}. 

It is important to keep in mind that properties of galaxy forming in a halo  and properties of its CGM may depend on the choices of our model parameters (e.g, the diffusion coefficient). The parameters have sizeable uncertainties and thus, in principle, they need to be varied to explore their full effect on the CGM \citep[e.g][]{Salem2014}. In this pilot study, we choose to investigate results only for the fiducial parameter values, chosen to be consistent with current observational constraints and theoretical expectations. 

Note also that CRs can lead to gamma ray emission via pion production. However, in our model, we find that only $<1\%$ of the total energy would be emitted as gamma rays, consistent with a detailed CR modelling for the Milky Way \citep{Strong2010}. Therefore, the gamma rays should not have a significant effect on the low-density circumgalactic medium, but can be used as interesting observational constraints on the cosmic ray feedback in the future.

%-------------------------
\section{CGM Analysis}
\label{sec:CGManalysis}
%-------------------------

\subsection{Lines of sight and ion column densities}

Observed quasar spectra reveal existence of a wide range of ion species tracing different temperatures, which indicates that the CGM is multi-phase.
It is thus important to use the entire range of ion species to test whether simulations reproduce the observed CGM structure.  

We compute abundances of all the ions for which extensive observational measurements have been reported in the literature. These ions cover a wide range of ionization energies, $E_{\rm ion}$, as shown in Table 2. We will adopt a convention to name species  with $E_{\rm ion} < 54.4$ eV as {\it low ions} (e.g., H\,I, Mg\,II, Si\,II, Si\,III, Si\,IV and C\,II), those with $54.4 < E_{\rm ion} < 100 $ eV as {\it intermediate ions\/} (C\,IV), and those with $E_{\rm ion} > 100$ eV as {\it high ions\/}  (O\,VI). 

 Following \cite{Smith2011} and  \cite{Hummels2013}, we assume that the number density of ion $j$ of an element $X$ can be computed as: 
\begin{equation} n_{X_j} (n_H, T, Z) = f_{X_j} (T, n_H) f_X(Z)  n_H,\end{equation}
where $f_X(Z) = n_X / n_H$ is the fraction of the element $X$  relative to hydrogen and $f_{X_j} (T, n_H) = n_{X_j} / n_X$ is the ionization fraction of X in state $j$. 

We compute an interpolation table of $f_{X_j} (T, n_H,J_{\rm UV} )$ using the \texttt{CLOUDY} code \citep[version 13.02; last reviewed in][]{Ferland2013} under assumption of ionization equilibrium for a grid of $n_H$, $T$, and redshift ($z$, corresponding to different cosmic mean $J_{\rm UV}(z)$). For $J_{\rm UV}$ we adopt the redshift dependent ionization background HM05 \citep{HaardtMadau2012}, which includes contribution from QSOs and galaxies and assume that gas is optically thin throughout the CGM.  We assume solar pattern of heavy element abundances \citep{Asplund2009} and solar metallicity of $Z_{\odot} = 0.02$, and approximate the mean molecular weight by a constant $\mu = 0.62$ for ionized gas with $T > 10^4$ K, and $\mu = 1$ otherwise. 

The column density of ion $X_j$ along a given line of sight through simulation volume is computed as: 
\begin{equation} N_{X_j} = \int_{L} n_{X_j}(n_H, T, Z,z) dl,  \end{equation}
where $n_{X_j}$ is 3D number density. 

To sample the CGM uniformly under all possible viewing angles, we uniformly sample impact parameter $d$ between the putative QSO and simulated galaxy and solid angle at the point in the line of sight (LOS) closest to the galaxy. To make sure that column densities for the LOSs near the edge of the box (i.e., large $d$) are not biased low, we have chosen a large box size $L_{\rm box} = 1$ Mpc (comoving) compared to the virial radius of the simulated dark matter halo. Furthermore, to ensure that column densities are not biased up to $d \sim 2.5 R_{\rm{vir}}$, 
we do not include the LOS with path lengths $ \apl 0.27 L_{\rm box}$ in our analysis and plots. 

\begin{figure}
\begin{center}
\includegraphics[scale=0.45]{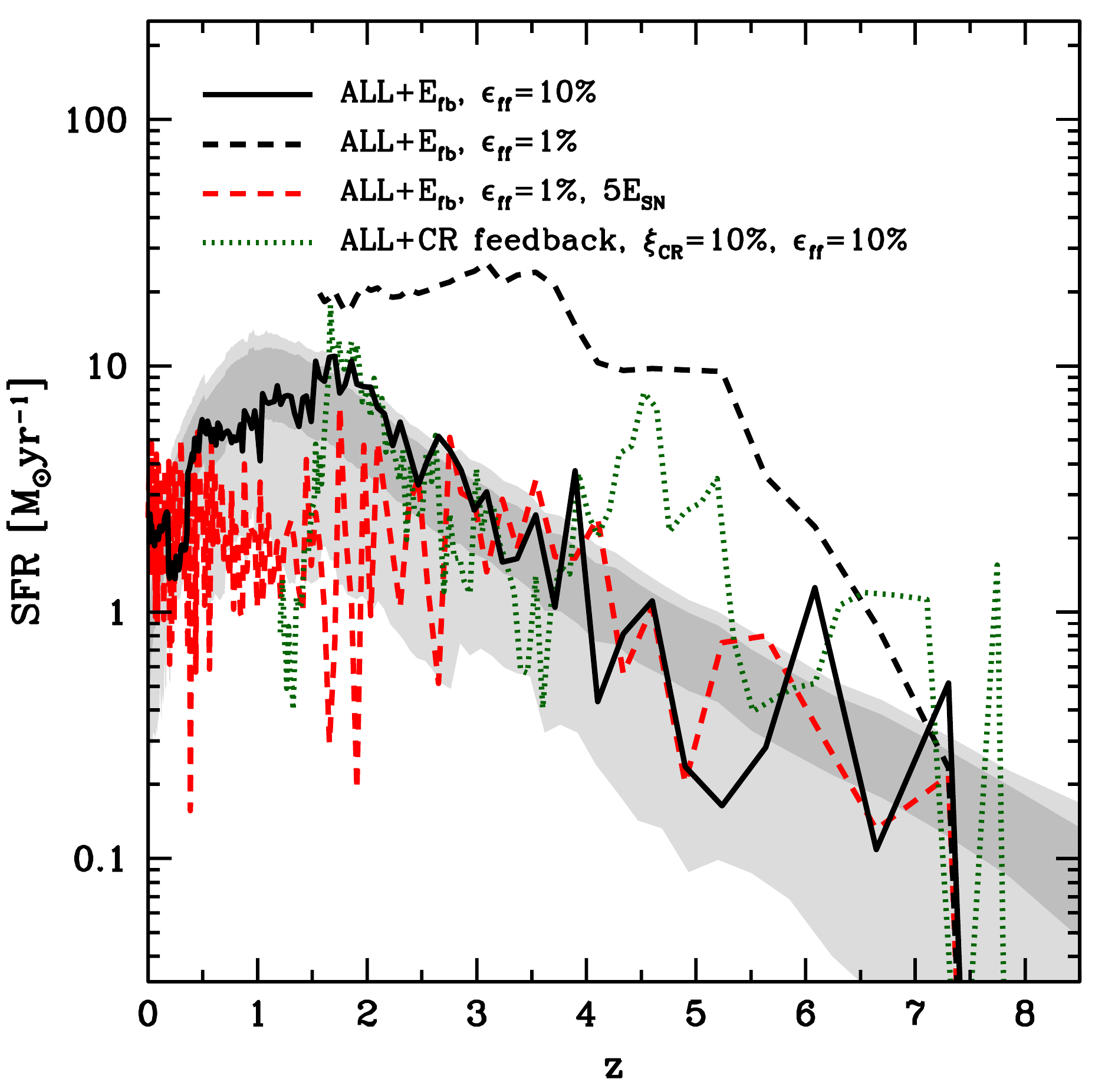}
\caption{Star formation histories of the main galaxy in our simulations compared to the average histories for observed galaxies in halos of $M_{\rm vir}(z = 0) = 10^{12} M_{\odot}$ derived by \protect\citet[][with dark and light gray shaded bands showing the one- two-sigma confidence regions of their constraints, respectively]{Behroozi2013}. The SFHs in simulations are averaged over time bins of $\Delta t_{\rm SF}$ = 100 Myr.  \label{fig:SFHcomp} }
\end{center}
\end{figure}

\begin{figure}
\begin{center}
\includegraphics[scale=0.45]{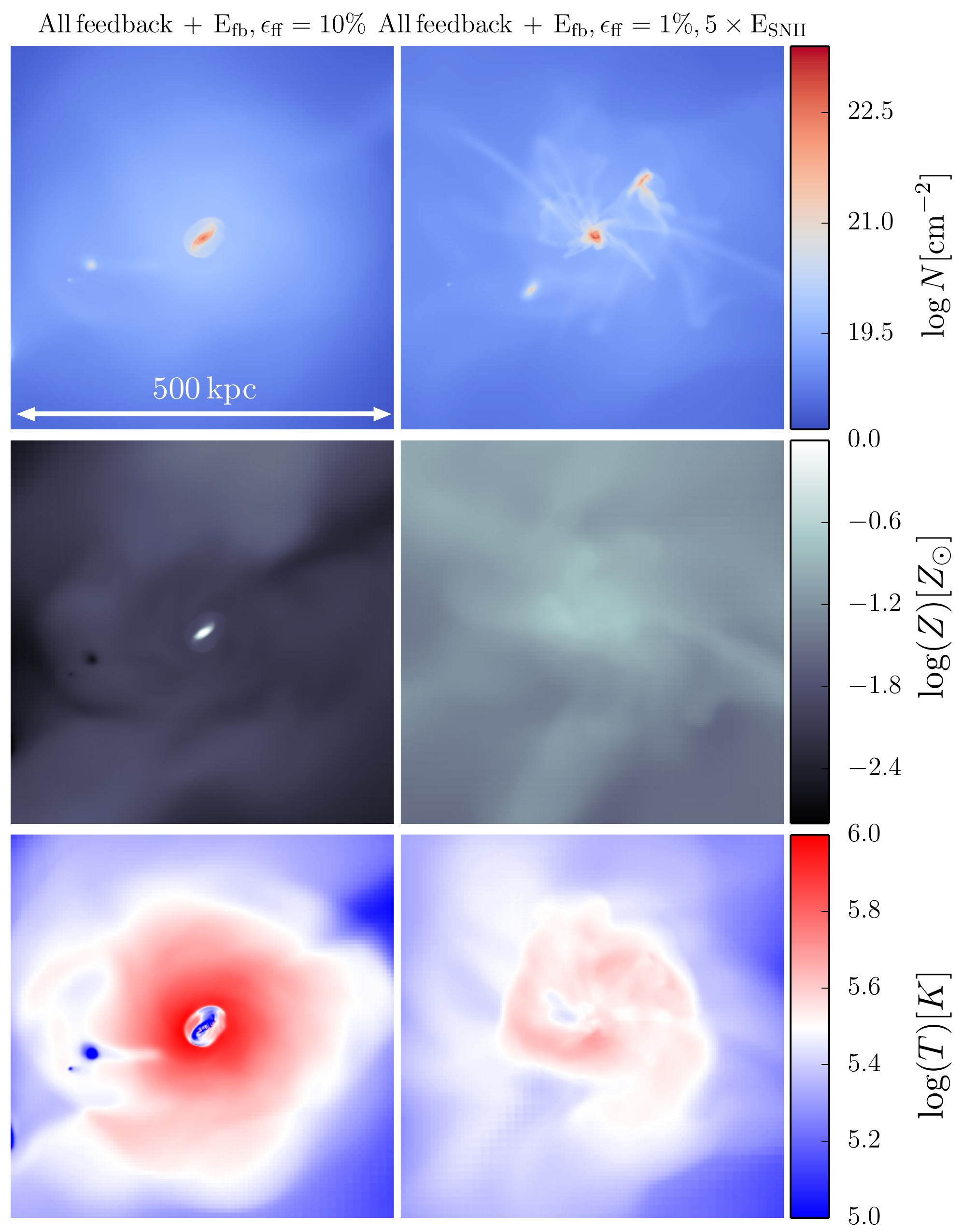}
\caption{Maps of the total gas column density, metallicity, and mass-weighted temperature of the fiducial and $5E_{\rm SNII}$ models at $z = 0$. The maps show the outflow streams aided by the boosted supernova events while the fiducial run maintains a normal star-forming disk. As a consequence, gas metallicity is well-mixed in $5E_{\rm SNII}$ model out to large radii at 10\% solar to solar value. In contrast, despite the disk in the fiducial run having solar metallicity, the gas hovers at a few percent solar in the circumgalactic space. Note that the color of the star-forming disk in the fiducial run has been clipped in order to ensure the dynamic range of the color map in the halo. \label{fig:galmaps}}
\end{center}
\end{figure}

\subsection{Curve of Growth Analysis}

In observational samples, direct column density measurements are often available only in relatively high signal-to-noise and high resolution spectra, while some studies only provide the equivalent width $W_r$ of absorbers. To combine all the data sets in a uniform manner, we follow \cite{Hummels2013} and apply a curve of growth analysis to convert from column density to equivalent width and vice versa using an approximation of equation (9.8) \& (9.27) in \cite{Draine2011}:

\begin{equation} 
W_{r} \approx \lambda \times \left\{
\begin{array}{lr}
\sqrt{\pi} \frac{b}{c} \frac{\tau_0}{1 + \tau_0 / (2 \sqrt{2})}\ \ \mathrm{for}\ \mbox{$\tau_0 <$ 1.254}\\
\left[\left(\frac{2b}{c}\right)^2 \ln{\left(\frac{\tau_0}{\ln{2}}\right)} + \frac{b\gamma_{lu} \lambda_{lu}}{c^2} \frac{(\tau_0 - 1.254)}{\sqrt{\pi}}\right]^{1/2}, \  \tau_0 > 1.254
\end{array}
\right.
\label{eq:equiwidth}
\end{equation} 
where
\begin{equation}
\tau_0 \approx \sqrt{\pi} \frac{e^2}{m_e c} \frac{N_l f_{\rm lu} \lambda_{\rm lu}}{b}.
\label{eq:optdepth}
\end{equation}  
$W_r$ is the rest frame equivalent width in units of wavelength, $\tau_0$ is the optical depth given the column density $N_l$  and $b$ parameter of the specie in state $l$, $\gamma_{\rm lu}$, $f_{\rm lu}$ and $\lambda_{\rm lu}$ are the intrinsic width, oscillator strength and wavelength of the specie transition from state $l$ to $u$. The atomic data for all transitions are taken from a compilation study by \cite{Morton2003}.

To compute $N$ from $W_r$, we invert the function $W_r(N,b)$ using its bivariate spline approximation. When only $W_r$ measurements are available, the set of $b$ values are determined by a Gaussian width $\sigma$, as $b \equiv \sqrt{2} \sigma$, which are taken from the literature \citep[e.g., see][]{Hummels2013, LiangChen2014}. To estimate
$W_r$ from $N$ in simulations, we treat $b$ as a free parameter, because simulations do not model non-thermal components of the gas. We discuss this in detail in section \ref{sec:simobs}.

\begin{figure*}
\begin{center}
\hspace{20mm}
\includegraphics[scale=0.5]{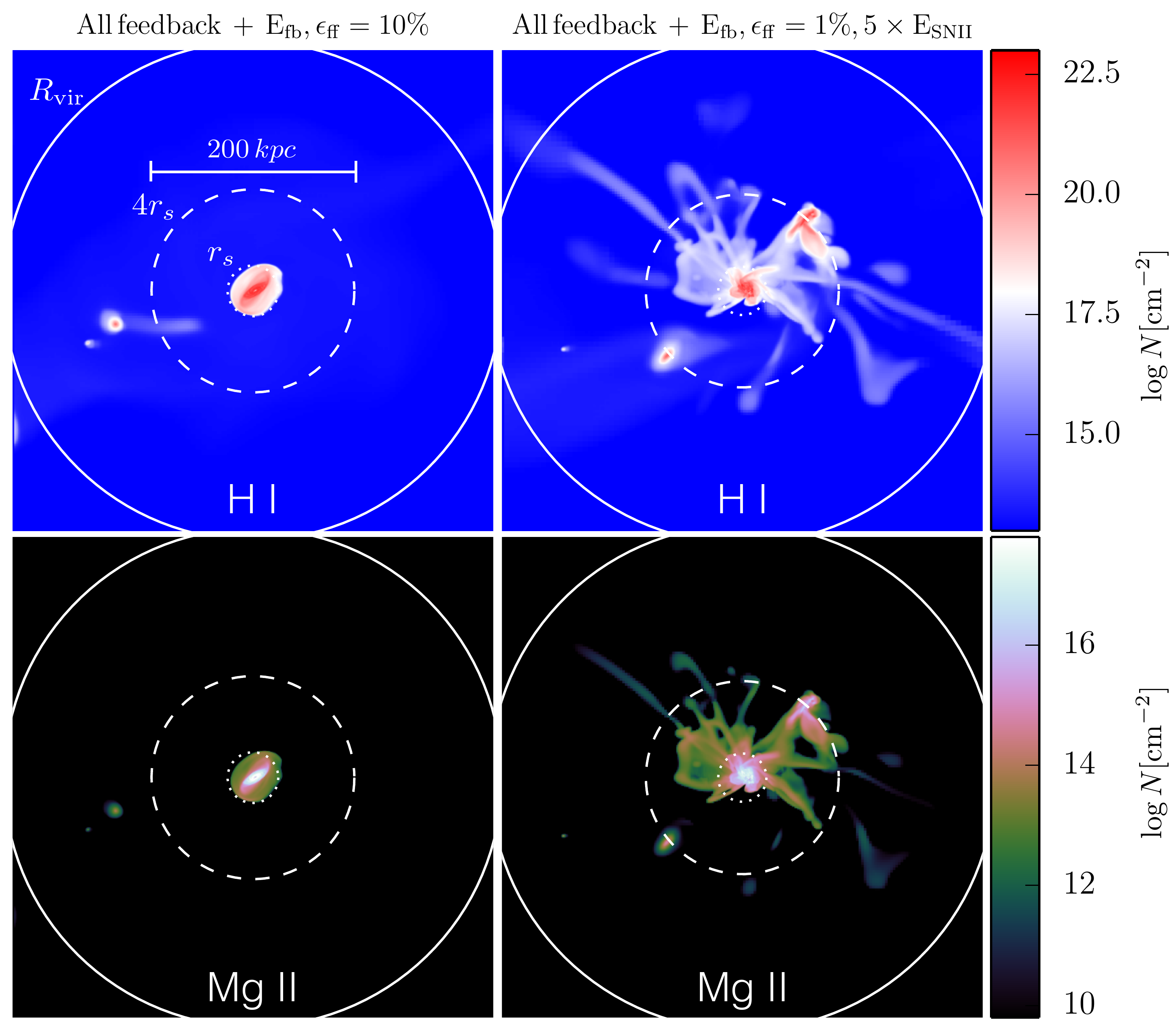}
\caption{The spatial column density distribution (500 kpc across; $R_{\rm{vir}} \approx 250$ kpc) of the chemically enriched CGM as probed by H\,I and Mg\,II at $z = 0.0$. Solid circle represents $R_{\rm{vir}}$, dashed circle is $4 r_{s}$ and dotted circle is $r_s$. Our fiducial model with star-formation efficiency $\epsilon_{\rm{ff}} = 10\%$ produces a stellar component consistent with observations \citep{AgertzKravtsov2015,AgertzKravtsov2016} but fails to produce an extended multi-phase gaseous halo. On the other hand, our strong feedback model (with five times supernova energy, $5 \times \rm{E_{SNII}}$) matches the CGM profiles despite destroying the star-forming disk. This shows that the CGM provides sensitive orthogonal constraints on galaxy formation, especially feedback recipes. \label{fig:Nmaplow}}
\end{center}
\end{figure*}

\begin{figure*}
\begin{center}
\hspace{20mm}
\includegraphics[scale=0.5]{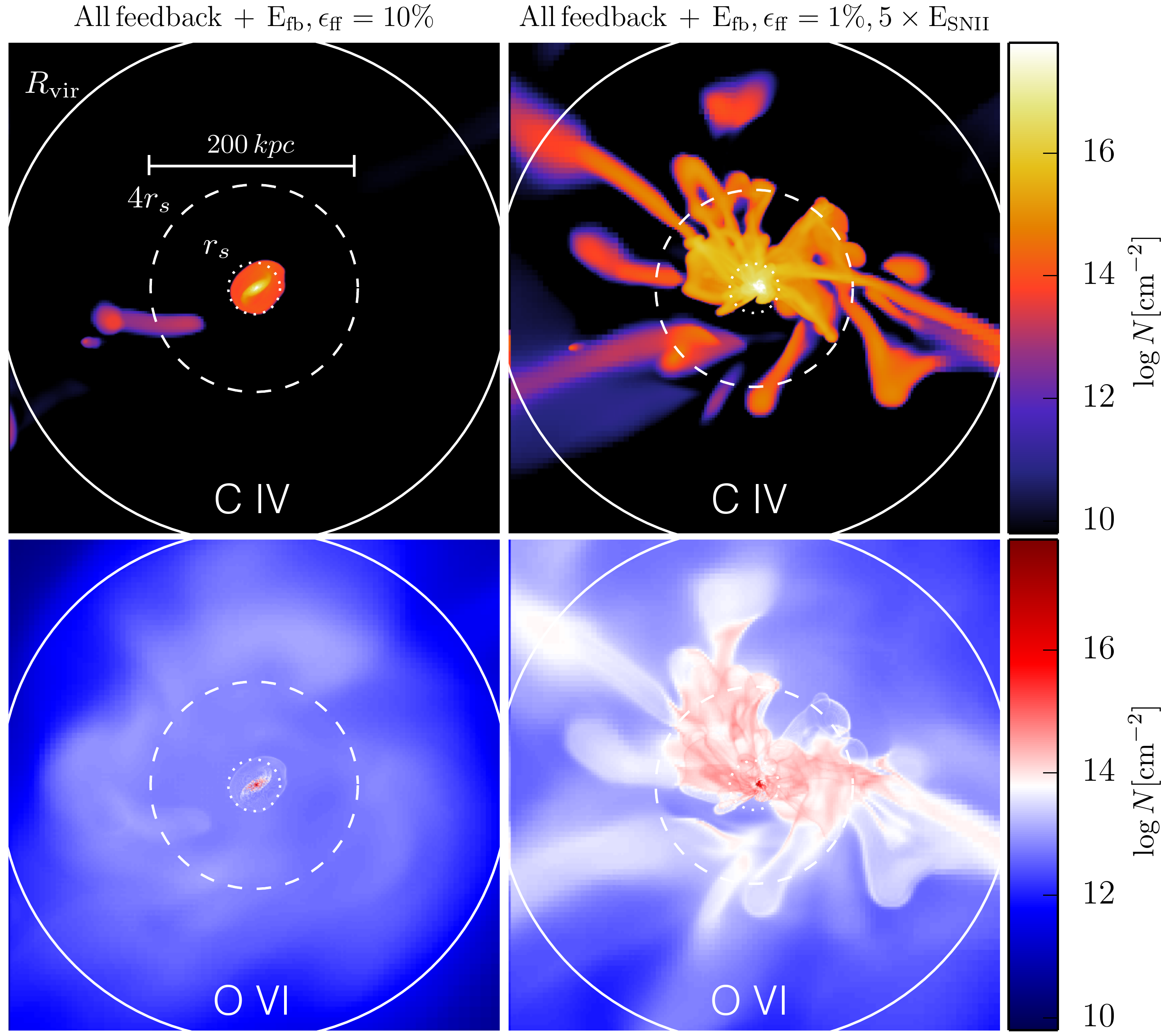}
\caption{Similar to Figure \ref{fig:Nmaplow}, but for intermediate ion C\,IV and high ion O\,VI at $z = 0.0$. The multi-phase gaseous halos are progressively more extended from low and intermediate ions (e.g., Mg\,II and C\,IV) to high ion (O\,VI).\label{fig:Nmapinterm}}
\end{center}
\end{figure*}

%---------------------------------------------------------
\section{CGM in Simulations}
\label{sec:cgmsim}
%---------------------------------------------------------
In this section we discuss the main qualitative features of the CGM in our simulated galaxies. We discuss their dependence on the feedback models (different implemented physics), as well as varying parameters of star formation and feedback recipes. 

Figure \ref{fig:SFHcomp} shows star formation histories of all the simulations used in this study compared to the range of star formation histories derived for observed galaxies hosted by halos of similar mass \citet{Behroozi2013}. We can see that the main progenitor in the {\tt ALL\_Efb\_e001} run  has star formation rates at $z>2$ that are up to an order of magnitude higher than the rate inferred by \citet{Behroozi2013}. This leads to an overestimate of stellar mass of the final object by a factor of two and creation of a dense central spheroid in the stellar distribution of the galaxy at low $z$. In other runs, feedback is sufficiently efficient to suppress star formation to the level consistent with the inference of \citet{Behroozi2013}. At lower $z$ the star formation rate in the fiducial run {\tt ALL\_Efb\_e010}  continues to track the empirical SFH, while the rates in the {\tt ALL\_Efb\_e001\_5ESN}  and {\tt ALL\_e010\_CR} runs fall below the gray band. As we discuss below, the bursty nature of star formation in the {\tt ALL\_Efb\_e001\_5ESN} simulation, however, allows it to drive significant outflows out of the central disk even at low redshifts, as can be seen in Figure \ref{fig:galmaps}.  The {\tt ALL\_e010\_CR} run tracks the SFH derived 
from observations quite well down to $z=1.5$, although the SFH starts to decrease sharply after this epoch. 

As shown by \citet{AgertzKravtsov2015,AgertzKravtsov2016}, the run {\tt ALL\_Efb\_e010} not only reproduces the observed star formation history (SFH), but produces a galaxy with realistic stellar mass-halo mass ratio, bulge-to-disk ratio, metallicity, and the Kennicutt-Schmidt relation. On the other hand, the {\tt ALL\_Efb\_e001\_5ESN} run, with less efficient star formation but more energetic SNe events, under-produces stellar mass at low $z$, and fails
to produce a disk-dominated galaxy. 

\subsection{Main features of the CGM in simulations}

 Figures \ref{fig:galmaps}, \ref{fig:Nmaplow}, and \ref{fig:Nmapinterm} show the overall thermodynamic properties of the CGM and column density maps of specific ions at $z = 0$. Here we focus on the runs {\tt ALL\_Efb\_e010} and  {\tt ALL\_Efb\_e001\_5ESN} and will compare results with other runs in the next subsection. 
The figures clearly show that  these two runs produce not only very different central galaxies, but very different CGMs. The CGM of the {\tt ALL\_Efb\_e010} run is hot, low-metallicity, and is almost devoid of gas with $T\sim {\rm few}\times 10^5$ K in its inner regions. In contrast, the CGM in the  {\tt ALL\_Efb\_e001\_5ESN} simulation does have such warm gas due to gas lifted out of the disk by recent outflows. These can be seen as ``tails" and ``streaks'' in the column density map.

 Figures \ref{fig:Nmaplow} and \ref{fig:Nmapinterm} also show that the radial distribution of the high ions is more extended compared to the low ions.  Overall, analyses of \citet{Ford2014} and \citet{Ford2015} show that higher ionization energy ions like, O\,VI, originate from the gas that was ejected from the disk with outflows at higher redshifts, while low-ionization energy ions originate in gas associated with recent outflows.

\begin{figure*}
\begin{center}
\includegraphics[scale=0.45]{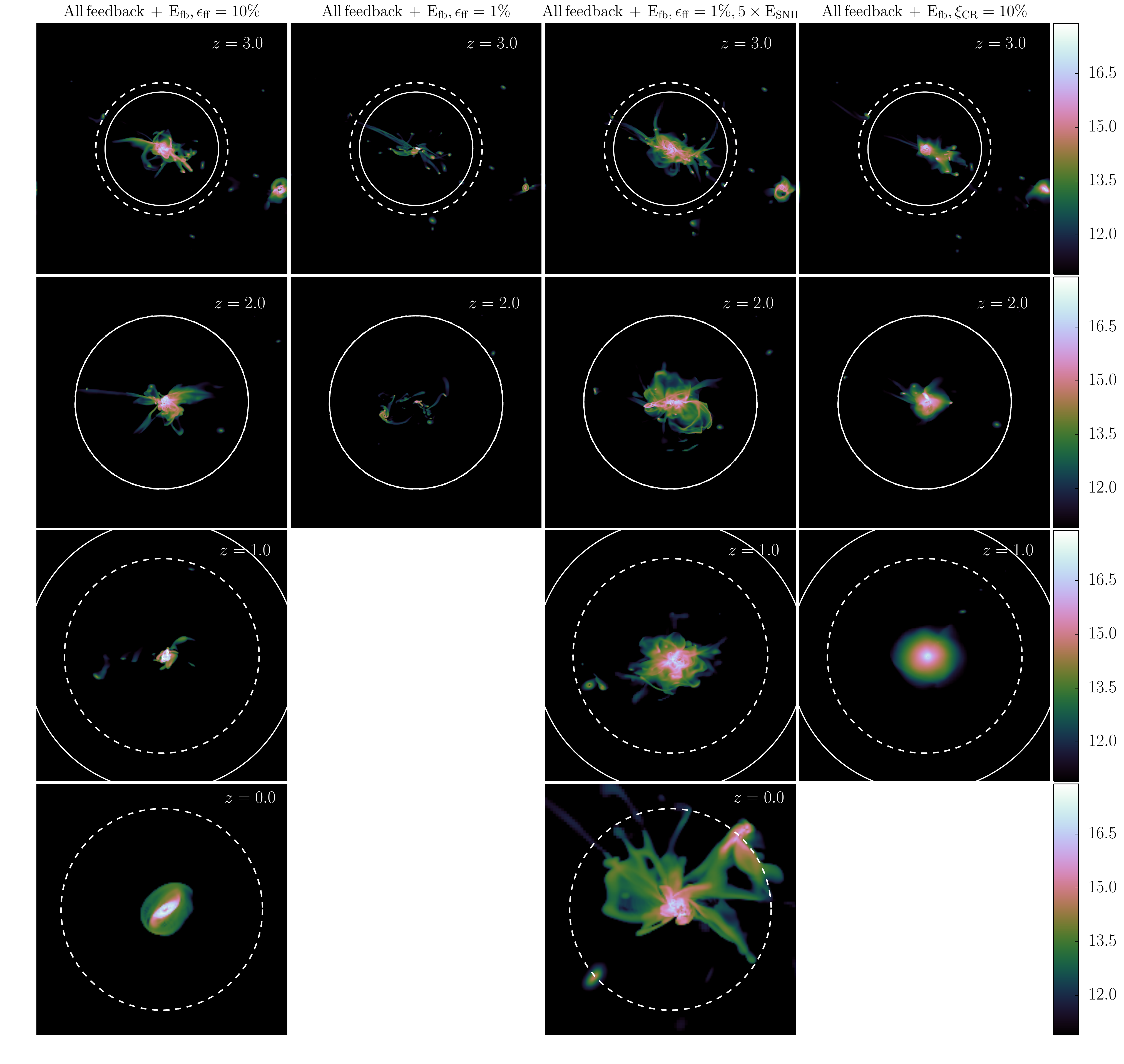}
\caption{ Model predictions of Mg\,II column density maps for different redshifts. The region is 250 kpc (physical) on a side. Solid circles represents $R_{\rm{vir}}$ (it is outside the region at $z=0$) and dashed circle represents 4$r_{s}$. At high redshifts, $ 4r_s$ tracks the growth of $R_{\rm vir}$, but at $z<2$ $R_{\rm vir}$ becomes larger and grows faster than $4 r_s$. For models that match observations of the CGM ({\tt ALL\_Efb\_e001\_5ESN} and {\tt ALLe010\_CR}) the physical extent of the CGM is approximately a constant fraction of $r_s$ at all $z$.  \label{fig:MgIIpro} }
\end{center}
\end{figure*}

\subsection{Cosmic Ray Feedback on the CGM}

\begin{figure*}
\begin{center}
\includegraphics[scale=0.81]{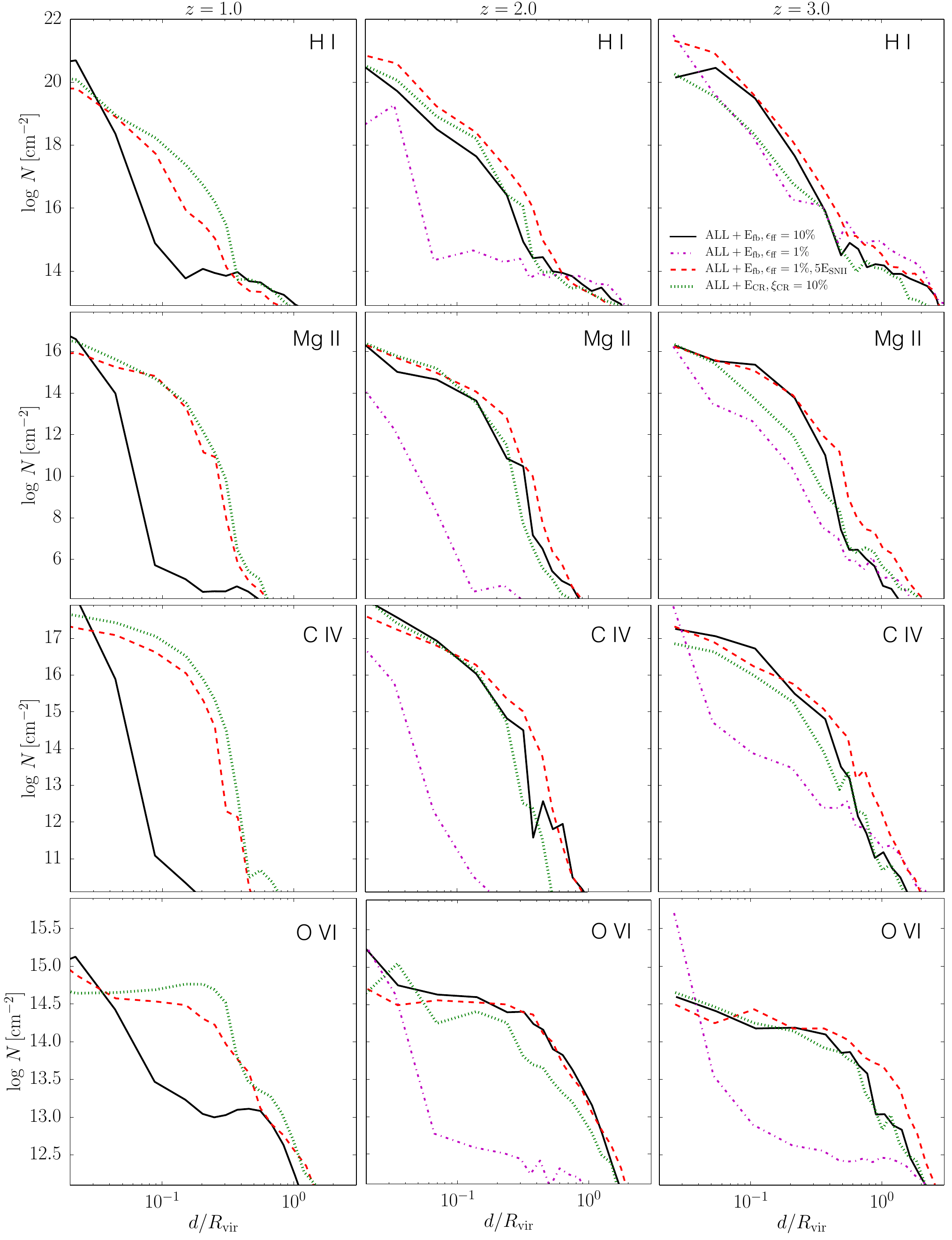}
\caption{Column density profiles for four representative low to high ions for our models and their evolution from $z = 3.0$ to $z= 1.0$. At high redshift ($z = 2- 3$), the CGM seems to be insensitive to the feedback recipes except for the very weak star formation efficiency $\epsilon_{\rm{ff}} = 1\%$. At $z = 1$, the fiducial model starts to deviate from $5 E_{\rm{SNII}}$ and cosmic ray models. The extent of the profiles in units of $R_{\rm vir}$ shrinks with redshifts mainly due to the fast-evolving $R_{\rm vir} $ relative to the physical extent. \label{fig:allpro}}
\end{center}
\end{figure*}

Recent studies \citep[e.g.,][]{Booth2013, SalemBryan2014} have shown that CR-driven winds can create large outflow mass loading factors, especially in dwarf galaxies. 
Figures \ref{fig:MgIIpro} and \ref{fig:allpro} show that both boosted SN-driven feedback in the run {\tt ALL\_Efb\_e001\_5ESN} and CR-driven outflows in the run {\tt ALLe010\_CR} produce column density profiles of similar shape and extent.    

However, the processes that  launch gas out into the CGM in these simulations are quite different. In the {\tt ALL\_Efb\_e001\_5ESN} run outflows are driven by energy and momentum injection at the sites of the SNe explosions in the disk and are quite bursty. In the CR run, on the other hand, the outflows are driven by a large-scale pressure gradient established by the cosmic rays diffusing out of the disk. The acceleration to  this gradient is more gradual, which results in lower outflow velocities and resulting lower gas temperature \citep[see][for more details on the wind properties in the CR feedback model]{Booth2013,SalemBryan2014,Salem2014}. The resulting distribution of the CGM in the CR run is much smoother than in the {\tt ALL\_Efb\_e001\_5ESN} run, as can be seen in the right column of Figure \ref{fig:MgIIpro}. 

\subsection{Evolution of the CGM in runs with different feedback models}

Figure \ref{fig:MgIIpro} also shows evolution of the CGM  in our different runs at different $z$. We choose to show  Mg\,II $\lambda \lambda 2796, 2803$ lines as an illustrative example, but similar behavior is seen for other ions that trace the cold/warm gas (e.g., C\,IV).  Figure \ref{fig:MgIIpro} shows that the extended gaseous halo is established by $z = 3$ in all runs except {\tt ALL\_Efb\_e001}. In the case of {\tt ALL\_Efb\_e001}, the gas in the halo collapses onto the central disk where it is converted into stars, which results in high star formation rates (see Fig. \ref{fig:SFHcomp}).  The ubiquity of the extended CGM in all runs at high $z$ is due to the shallow gravitational potential at early epochs, which allows even moderate feedback to eject gas from the disk and launch it far into the halo (and even beyond). It is thus more difficult to use CGM for differentiating feedback recipes at high redshift ($z \ge 2 - 3$).

 At redshifts $z \le 2$, the galaxy in the simulation {\tt ALL\_Efb\_e010} continues to form stars with the rate that matches the empirical average SFH of \citet{Behroozi2013} for galaxies of this mass. However, the feedback implementation fails to drive winds far into the halo at $z<1$; the SN action at these redshifts is limited to small fountains and stirring of the gas within the disk. On the other hand, the runs {\tt ALL\_Efb\_e001\_5ESN} and {\tt ALLe\_010\_CR} drive significant outflows down to the smallest redshifts to which they were run. The winds lift cool gas from the disk into the halo in extended plumes, which are prominent in Figures \ref{fig:Nmaplow}--\ref{fig:MgIIpro}. 

Another interesting feature of the evolution shown in Figure \ref{fig:MgIIpro} is that the extent of the high-column density area increases visibly slower than the virial radius (shown by the solid lines). The same can be seen in the column density profiles at different $z$ in Figure \ref{fig:allpro}, in which profiles are rescaled by host halo $R_{\rm vir}(z)$ at the corresponding redshift. For example, although visible physical extent of the Mg\,II distribution increases from $z=3$ to $z=1$ in runs {\tt ALL\_Efb\_e001\_5ESN} and {\tt ALL\_e010\_CR} in Figure  \ref{fig:MgIIpro}, the extent in units of virial radius in Figure \ref{fig:allpro} decreases over the same redshift interval. 

The virial radius of host halos has now been used in a number of studies to rescale the input parameter of galaxies at different redshifts in order to compare their profiles \citep{Churchill2013, Werk2014, LiangChen2014, Bordoloi2014}. However, the evolution of the profile in simulations indicates that the actual
scaling may be different.  This motivates a closer look at the evolution of the column density distribution and corresponding scaling across redshifts.

\begin{figure*}
\begin{center}
\includegraphics[scale=0.81]{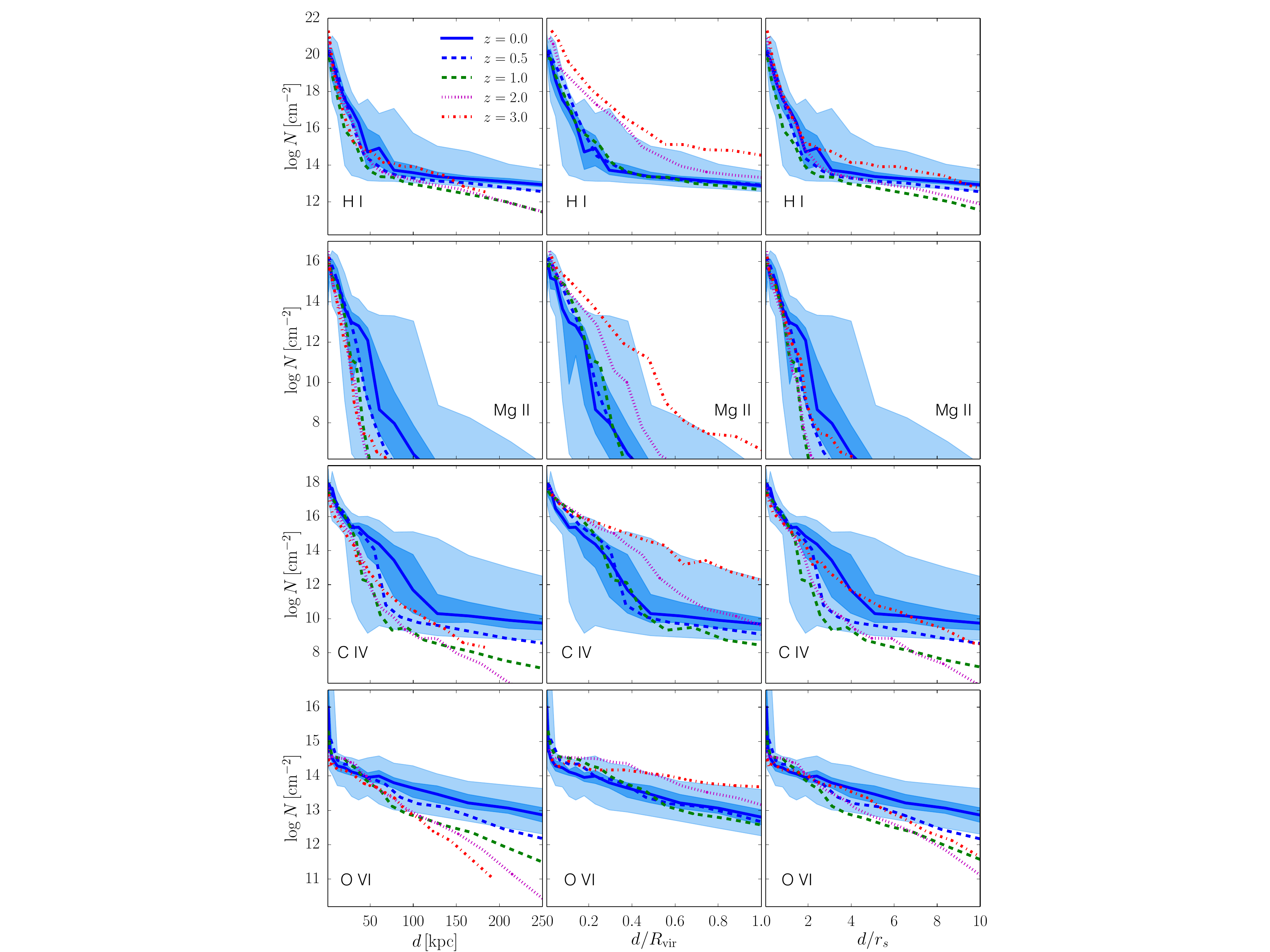}

\caption{Column density profiles from low (H\,I and Mg\,II) to high (C\,IV and O\,VI) ions in the {\tt ALL\_Efb\_e001\_5ESN} run. Various lines are medians of the profiles at different redshifts (blue to red as low $z$ to high $z$). Dark and light blue shaded regions represent 68\% and 95\% intervals for the profiles at $z = 0$, respectively. The left column shows the column density profiles in the inner regions remained relatively unchanged over 11 billion years at $z\leq 3$. This is why scaling with slower evolving scale radius, $r_s$ (right panel) brings profiles at different $z$ into agreement better than scaling with $R_{\rm vir}$ (middle column). At the same time, scaling with $r_s$ also accounts for the different masses of the galaxies, which is necessary to put a population of observed galaxies in an equal footing. Note that higher ionization ions have progressively more extended profiles.
 \label{fig:proevol}}
\end{center}
\end{figure*}

\subsection{Evolution of the radial distribution of the CGM column density and rescaling with halo scale radius}
\label{sec:simproevol}

The boundary of dark matter halos is usually defined as the radius enclosing a given density contrast $\Delta$ (e.g., $\Delta$= 200) relative to  a reference density, such as the critical, $\rho_{\rm crit}(z)$, or mean, $\rho_{\rm m}(z)$, density of the universe. Common choices are $R_{\rm vir}$ (adopted in this paper), $R_{\rm 200m}$ and $R_{\rm 200c}$. These definitions of the radii are loosely motivated by simple models of spherical top-hat collapse. For real halos the increase of the virial radius roughly tracks the actual physical splashback radius $R_{\rm sp}$ \citep{more2015}, which delineates the outer region enclosing recently accreted matter.
During epochs when halos accrete quickly, the entire halo profile scales well with the virial and splashback radii.  However, when the accretion rate slows down the inner density profile remains almost static in physical units \citep[e.g.,][]{Cuesta2008} and thus no longer scales well with the outer virial or splashback radius. Galaxies at $z\leq 2$ are in the latter regime, which means that scaling with the virial radius may not be optimal, especially since observations probe the CGM mostly in the inner halo ($\le 4-6 r_s$).

Evolution of the CGM for run {\tt ALL\_Efb\_e001\_5ESN}  is illustrated in Figure \ref{fig:proevol}, which shows evolution of the column density profiles of different ions from $z =3$ to $z = 0$  in physical kpc and scaled with $R_{\rm vir}$ in the left and middle columns.  Over the range of  redshifts shown the dark matter halo mass grows from $M_{\rm vir} = 3 \times 10^{11} M_{\odot} $ to $M_{\rm vir} = 10^{12} M_{\odot} $, while $R_{\rm vir}$ grows by by a factor of 4.6.  The figure shows that the profile in physical kpc evolves only very weakly over 11 billion years since $ z= 3$ and that scaling with virial radius introduces spurious evolution. 

\begin{figure}
\begin{center}
\includegraphics[scale=0.55]{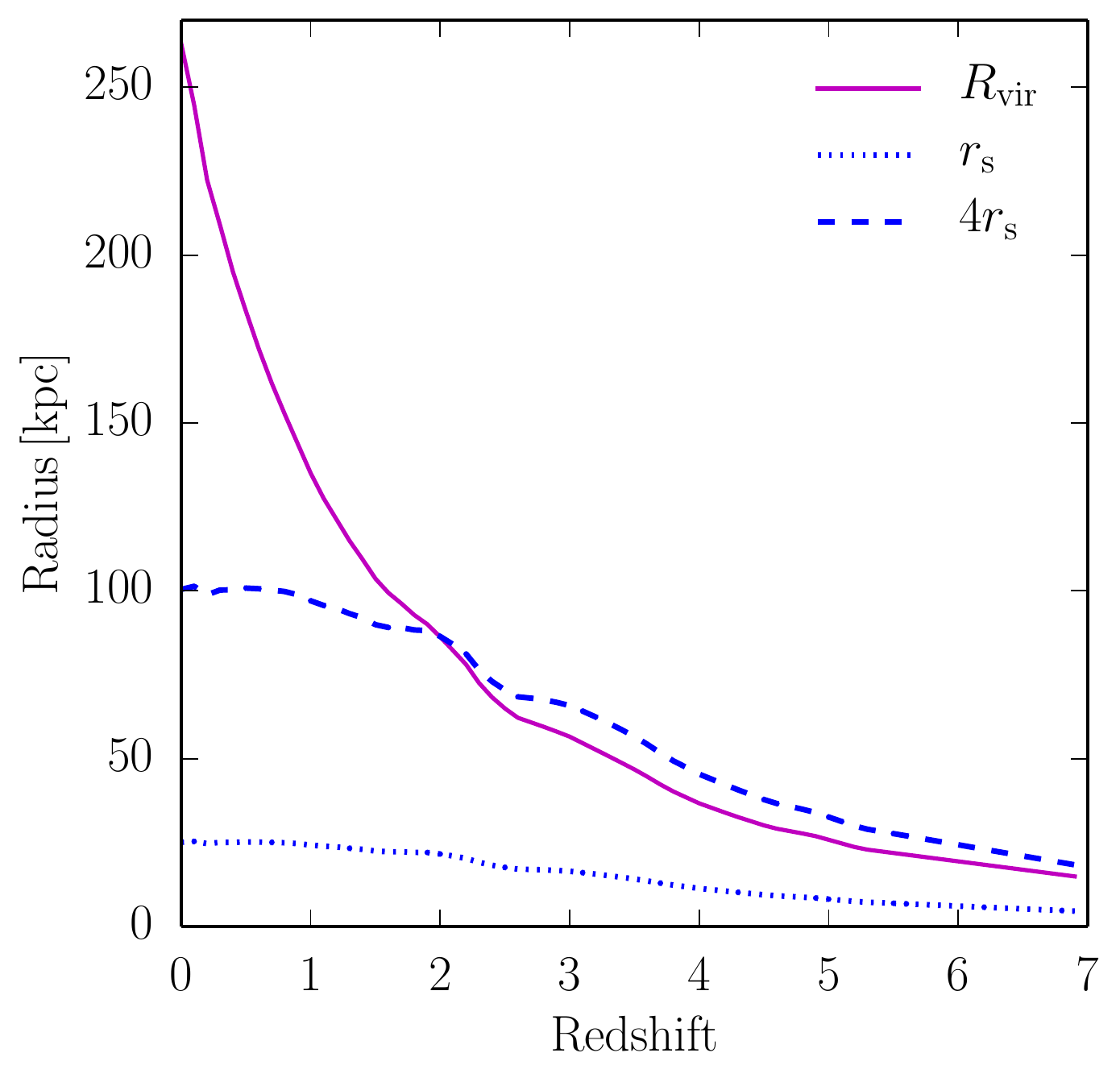}
\caption{Evolution of the virial radius and halo scale radius, $r_s$. Virial radius is measured from the dark matter halo at each snapshot of the simulation \texttt{ALL\_Efb\_e001\_5ESN}. The corresponding scale radius $r_s \equiv R_{\rm vir} / c_{\rm vir}$ is calculated via the $M_{\rm vir} - c_{\rm vir}$ model in \protect\cite{DiemerKravtsov2015}. At high redshifts, the inner halo characterized by $4r_s$ tracks the growth of the halo virial radius, $R_{\rm vir}$, until $z \sim 2$. Afterwards, $R_{\rm vir}$ continues to grow, while the inner region of the dark matter halo approaches a constant value. \label{fig:rsrvir}}
\end{center}
\end{figure}

Recently, \citet{more2015} advocated a mass and radius definition to characterize evolution of the inner regions of halo profiles based on halo scale radius $r_s$, defined as the radius where logarithmic slope of the dark matter density profile is $-2$. In particular, they argued that the virial radius in the fast accretion regime is approximately equal to $4r_s$, while in the slow accretion regime the $r_s$ and $4r_s$ cease evolution \citep{Bullock2001} while $R_{\rm vir}$ continues to grow due to pseudo-evolution \citep{Diemer2013}, as can be seen in Figure \ref{fig:rsrvir}. 

If the gravitational potential well in the inner region plays a major role in controlling the extent of the CGM, it is natural to assume that the CGM profiles should scale with $r_s$. Indeed, the right column of Figure 
\ref{fig:proevol} shows that the column density profiles scale
with $r_s$ much better than with $R_{\rm vir}$.\footnote{ To estimate the scale radius $r_s\equiv R_{\rm vir}/c_{\rm vir}$, we use the halo-concentration model of \cite{DiemerKravtsov2015}, implemented in the publicly available python code {\tt Colossus}: {\tt http://bdiemer.bitbucket.org}}

Note that the scaling of profiles around galaxies of different stellar masses at a given $z$ with $r_s$ (or any multiple of it) is very similar
to scaling with $R_{\rm vir}$ at the same $z$ because $r_s=R_{\rm vir}/c_{\rm vir}$ and $c_{\rm vir}\propto M_{\rm vir}^{-0.1}$ is a weak function of mass. However, both $R_{\rm vir}$ and $c_{\rm vir}$ evolve fast with redshift in a way that leaves $r_s$ almost constant at $z<2$ for galaxy sized halos. Thus, re-scaling of profiles of objects at different $z$ with $r_s$ and $R_{\rm vir}$ is quite different. 

Although concentration at a fixed halo mass exhibits scatter of $\approx 0.14$ dex, which will be added in quadrature to the scatter in the 
$M_*$ at fixed $M_{\rm vir}$ during conversion from $M_*$ to $M_{\rm vir}$, Figure  \ref{fig:proevol} shows that such additional scatter may be worth the reduced bias in the redshift rescaling. As we will show in the next section, observed column density profiles also appear to favor 
rescaling with $r_s$. 

\subsection{Exponential form of the column density profile and scaling with ionization energy of ions}
\label{sec:exppro}

The column density profiles plotted on the log-log scale in Figure \ref{fig:allpro} exhibit a sharp turnover at a particular impact parameter. Although often interpreted as indication of an ``edge'' to the ion distribution, such sharp turnover can also simply reflect the smooth exponential decrease of column density. 
Indeed, the same profiles of the \texttt{ALL\_Efb\_e001\_5ESN} run shown on the log-linear scale in Figure \ref{fig:proevol} can clearly be well approximated by an exponential profile (i.e., a straight line in the log-linear plot). Some ions exhibit two exponential components, but the outer component is at the column densities that are well below the sensitivity of current observations. 

The exponential profile $N(d) = N_0 \exp (-d/h_s)$, or equivalently $\log_{10} N(d) = \log_{10} N_0 - \frac{d}{h_s * \ln (10)}$, is characterized by a single scale, which we will call the {\it scale height}, $h_s$, to differentiate from the exponential scale {\it lengths} of galaxy light distribution. 

The scale height of the profile provides a simple estimate of the extent of the CGM. \cite{Nielsen2013} and \cite{Borthakur2015} show that the observed column density profiles of some of the ions can indeed be described by an exponential profile. Note, however, that in both studies the exponential distributions are fitted to the equivalent width profiles instead of the column density. Due to non-linearity of the conversion between $N$ and $W_r$, scale heights from our fits should not be \textit{directly} compared with ones from $W_r$ profiles. Instead, a conversion from $N$ to $W_r$  is needed before comparisons (see Figure \ref{fig:Wrevol}). 

\begin{figure}
\begin{center}
\hspace{8mm}
\includegraphics[scale=0.59]{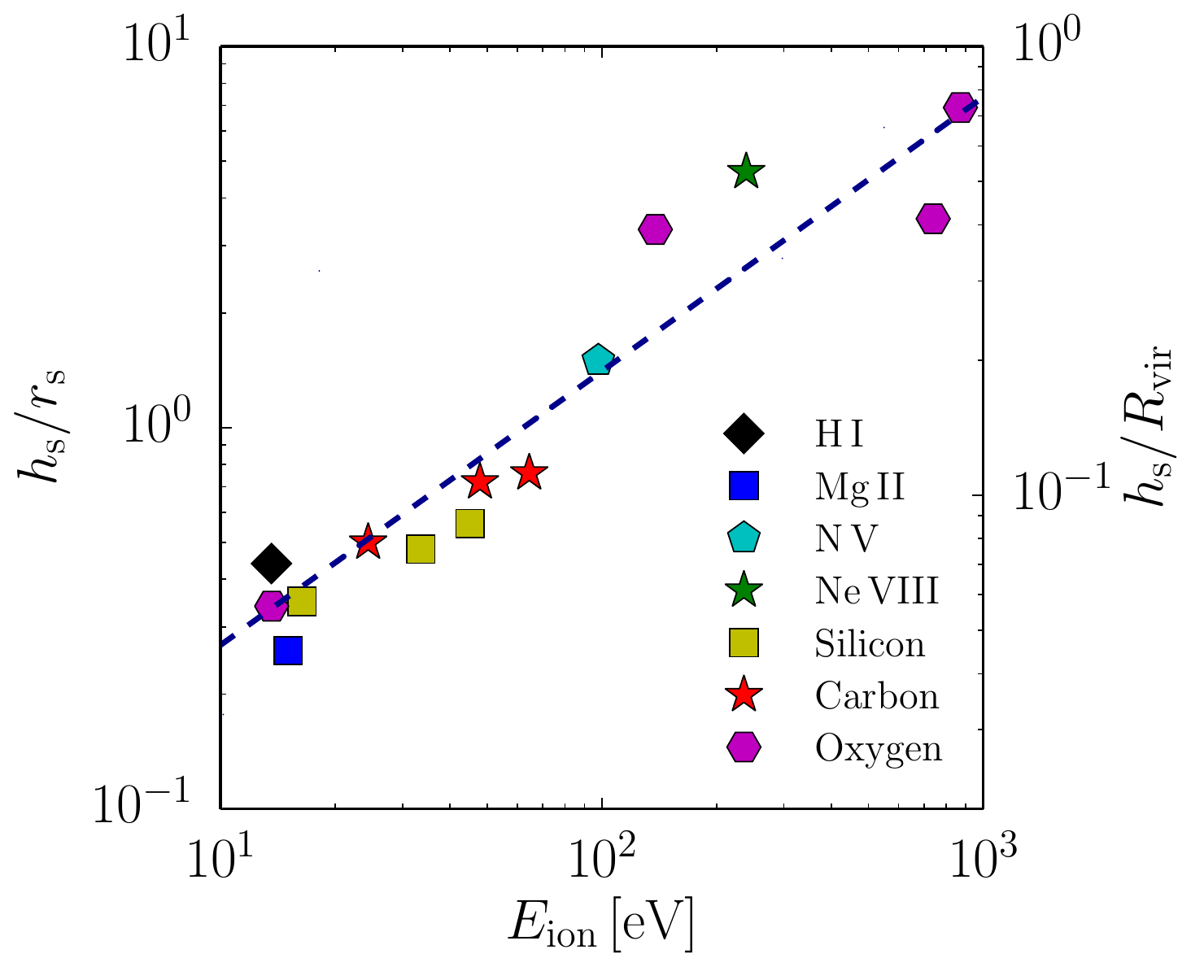}
\caption{Scale height of the exponential profile of the column density as a function of the impact parameter predicted in our run {\tt ALL\_Efb\_e001\_5ESN} at $z$ = 0 as a function of ionization energy of the corresponding ion. The left axis shows the scale height normalized to the halo scale  radius, while the right axis shows the scale height normalized to the halo virial radius. Color points show results of the fits of the column density profile of a particular ion, while the dashed lines show the best fit power law relations (see eqs. \ref{eq:mbfitvir}-\ref{eq:mbfitrs}). \label{fig:hEion}}
\end{center}
\end{figure}

The best fit scale heights of the run \texttt{ALL\_Efb\_e001\_5ESN} for the inner exponential profiles of different ions are shown in Fig. \ref{fig:hEion} and presented in Table \ref{tab:Eion}.
Interestingly, Figure \ref{fig:hEion} shows that the scale height exhibits a tight power law scaling withion ionization energy: $h_s \propto E_{\rm ion}^m$. 
Previous theoretical \citep[e.g.,][]{Hummels2013, Ford2015} and observational \citep{LiangChen2014,Johnson2015} studies have also found that higher ions have more extended distributions compared to the low ions. Here we present the first quantitative characterization of this trend and demonstrate that it is a remarkably tight function of the ionization energy of ions. 

 The best fit power law relation between scale height normalization to the virial radius, $R_{\rm vir}$, or scale radius, $r_s$, is characterized by the slope $m$ and offset $b$: $\log_{10}(x)= m\log_{10}E_{\rm ion}+b$, where
\begin{equation}
\mathrm{for}\ x = h_s/r_s:\ \ m = 0.742 \pm 0.005,\  b_{r_s} = -1.29 \pm 0.02.
\label{eq:mbfitvir}
\end{equation}
\begin{equation}
\mathrm{for}\ x = h_s/R_{\rm vir}:\ \ m = 0.742\pm 0.005,\  b_{R_{\rm vir}} = b_{r_s} - \log c_{\rm vir}
\label{eq:mbfitrs}
\end{equation}
where $c_{\rm vir} \approx 10.5$ is the concentration parameter of the halo at $z = 0$, computed from the $M_{\rm vir} - c_{\rm vir}$ relation \citep{DiemerKravtsov2015}. We also find a similar power law slope $m = 0.72$ for the \texttt{ALL\_e10\_CR} at $z = 1$. We estimated the best-fit scale heights $h_s$ for each ion by a simple $\chi^2$ minimization fit to the simulated scaled profiles $\log N_{\rm ion} - d/ r_{\rm s}$.   The fit is done to the range of radii, $0.1 < d/r_{\rm s} < 4$, and column densities  $\log N_c < \log N < 17.2 \,\rm{cm^{-2}}$ (21 for H\,I), where $\log N_c$ is the column density where we can clearly see the transition between the inner and outer exponential distributions. The limits are chosen to exclude the ISM of the galaxy and limit the fit only to the inner exponential component. 

The errors on the parameters are estimated by bootstrapping the column density measurements of many lines of sight and thus account for
the scatter in the profiles.  We also varied the selected ranges of the data for the fit and find that $h_s$ can shift systematically by $\sim 5 - 10\%$. Nevertheless, the power law dependence of $h_s$ on $E_{\rm ion}$ remains the same. We therefore caution readers that when comparing scale heights with observations, the data should be fitted in the same range.

The trend of the scale height with ionization energy is likely due to two separate reasons. First, abundance of low ions is determined by photoionization equilibrium and their distribution traces a particular range of the ionization factor, $U$, and thus a particular range of gas density profiles \citep[see, e.g., \S 16.9 in][]{Mo2010}. Ions with higher ionization energies, such as C\,IV and O\,VI, can be created both by photo- and collisional ionization. Overall, they favor regions with higher $U$ and thus lower gas densities, which can be partly responsible for the trend we observe. The dependence of scale-height on the ionization energy of ions may thus be related to the fact outflowing plumes of gas expand as they move out to regions of lower density and pressure. 

To test this idea, we carried out an analysis with and without an ionizing background to compare the relative contribution from photo- and collisional ionization. Indeed, with the addition of photo-ionization, low ions are lifted into higher states in the low density regime (at large distances). This effectively boosts the scale heights of high ions, while decreasing the scale height of  the low ions. Thus, the ionization factor dependence on density is at least partially responsible for the $h_s-E_{\rm ion}$ scaling. 

The fact that higher ions arise from shocked gas that was carried out by older winds \citep[e.g.,][]{Ford2015} may also contribute to the correlation. 
An NFW-like potential  is quite shallow in terms of its logarithmic slope out to $r\approx 1r_s$  and steepens at large radii.  It is thus tempting to associate the characteristic scale height of the low ions ($\approx 2r_s$) with the region of the shallow potential out to which recent winds propagate easily and then stop when the derivative of the potential steepens. However, we have checked the evolution of gas outflows in our simulations and found that even the recent outflows in the \texttt{ALL\_Efb\_e001\_5ESN} run propagate to considerably larger radii than $r_s$. The scale height of low ions is thus likely determined by the overall shape of the gaseous halo density profile and the corresponding profile of the ionization factor.

The \textit{prediction} of the exponential profile and the tight correlation of the scale height with $E_{\rm ion}$ can be tested against observations. The correct treatment of the profile fitting in the observational data requires proper consideration of the upper limits and is out of the scope of our paper, but we plan to return to it in a future study.

%%%%%%%%%%%%%%%%%%%%%%%%%%%%%%%%%%%%%%%%%%%%%%%%
\begin{footnotesize}
\begin{table}
\begin{center}
 \footnotesize
\centering
%    \hspace*{-1.85 cm}
\begin{minipage}{100mm}
   \caption{Scale Heights From Simulations}
   \begin{tabular}{@{}ccccr@{}}
     \hline
     \hline
     \multicolumn{1}{c}{} Ionization State & $E_{\rm{Ion}}$ \lbrack eV\rbrack$^a$ & $h_s / R_{\rm vir} $ & $h_s / r_{\rm s}$\\
     \hline
     H\,I 	& 13.60 	& 0.042 $\pm$ 0.004 &       0.44 $\pm$ 0.04 \\
     
     \hline
     Mg\,II     & 15.04   & 0.025 $\pm$ 0.001	&   0.26 $\pm$ 0.01	\\

     \hline
     N\,V       & 97.89   & 0.142 $\pm$ 0.012	&   1.50 $\pm$ 0.13	\\

     \hline
     Ne\,VIII   & 239.09   & 0.447 $\pm$ 0.006 &    4.69 $\pm$ 0.07	\\

     \hline
     C\,II      & 24.38   & 0.047 $\pm$ 0.003	&      0.50 $\pm$ 0.03  \\
     C\,III      & 47.89   & 0.068 $\pm$ 0.003	&    0.72  $\pm$ 0.04 \\
     C\,IV      & 64.49   & 0.073 $\pm$ 0.004	&  0.76 $\pm$ 0.04 \\
     
     \hline
     Si\,II     & 16.35   & 0.033 $\pm$ 0.002	&  0.35 $\pm$ 0.02	 \\
     Si\,III    & 33.49   & 0.046 $\pm$ 0.002	&  0.48 $\pm$ 0.02 \\
     Si\,IV     & 45.14   & 0.053 $\pm$ 0.003 	&  0.56 $\pm$ 0.03 \\
     
     \hline
     O\,I      & 13.62      & 0.032 $\pm$ 0.002	&  	0.34		$\pm$ 0.02	\\
     O\,VI    & 138.12    & 0.315 $\pm$ 0.005	& 	 3.31  	$\pm$ 0.05	\\
     O\,VII   & 739.31    & 0.336 $\pm$ 0.004 	& 	 3.53	$\pm$ 0.04	\\
     O\,VIII 	& 871.38 	  & 0.659 $\pm$ 0.012	&  	6.91	$\pm$ 0.13	\\    
          \hline
	\multicolumn{4}{l}{$^a$Ionization energy from a compilation of atomic data in \protect\cite{Morton2003}.}\\
	\multicolumn{4}{l}{Scale heights $h_s$ from exponential fits, $\log_{10} N_{\rm ion} =\log_{10} N_0 -d/(h_s \ln (10))$,}\\
	\multicolumn{4}{l}{of the first component of the column density profiles for \texttt{ALL\_Efb\_e001\_5ESN}}\\
	\multicolumn{4}{l}{at $z = 0$. We see a power law dependence $h_s \propto E_{\rm ion}^{0.74}$. Note that } \\
	\multicolumn{4}{l}{$h_s / r_s = c_{\rm vir} \times h_s / R_{\rm vir}  $, where $ c_{\rm vir} \approx$ 10.5 is the concentration }\\
	\multicolumn{4}{l}{parameter, computed via $M_{\rm vir} - c_{\rm vir}$ model in \protect\cite{DiemerKravtsov2015}.} \\
	\multicolumn{4}{l}{Uncertainties on $h_s$ are computed via bootstrap sampling.}\\

 \label{tab:Eion}

  \end{tabular}
\end{minipage}
\end{center}
\end{table}
\end{footnotesize}
%%%%%%%%%%%%%%%%%%%%%%%%%%%%%%%%%%%%%%%%%%%%%%%%

%----------------------------------------------
\section{Radial column density distributions: simulations vs. observations}
\label{sec:simobs}
%----------------------------------------------

\begin{figure}
\begin{center}
\includegraphics[scale=0.65]{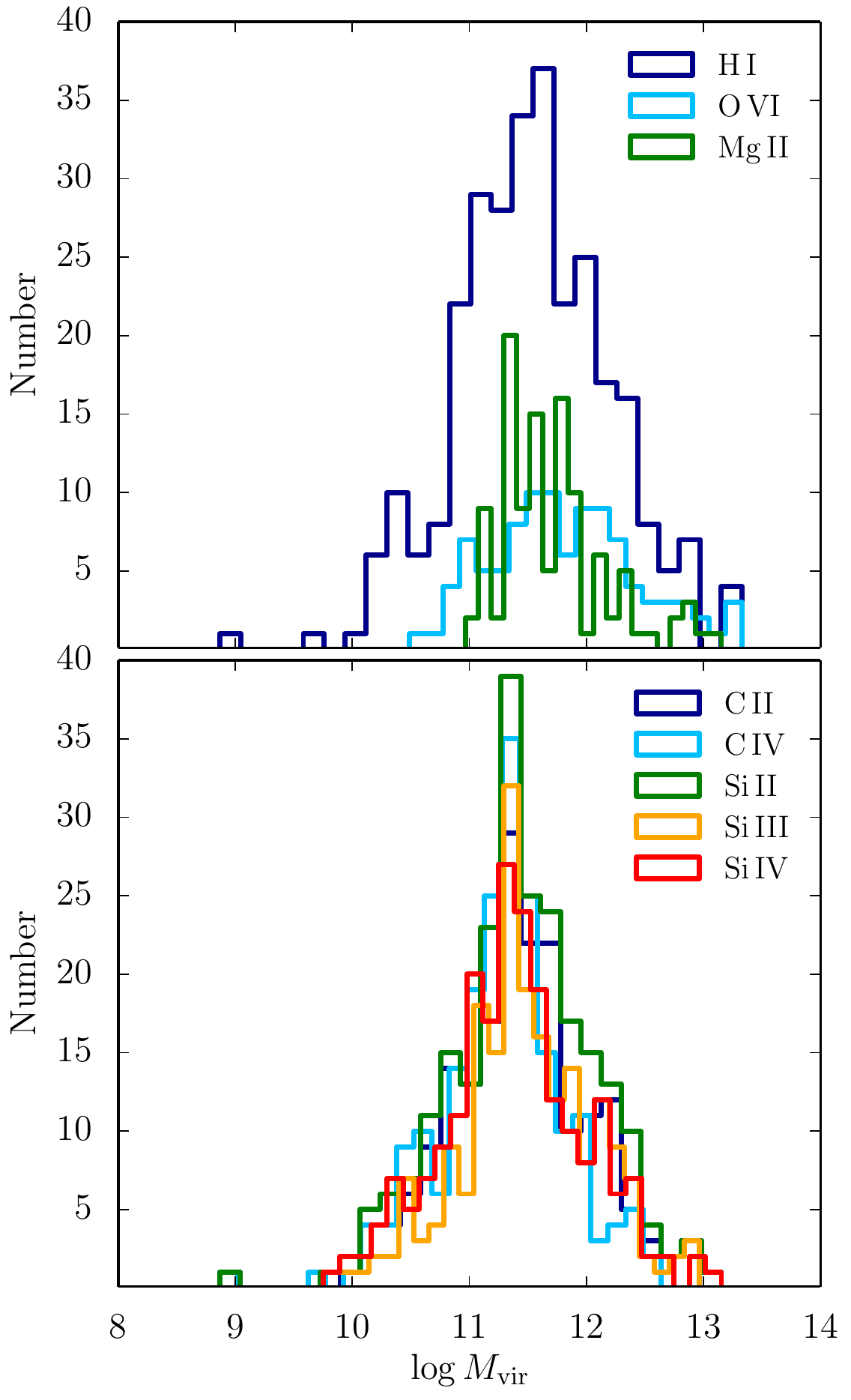}
\caption{Distribution of halo mass $M_{\rm vir}$ probed by various ions in the compiled low redshift data set.  See references in Table \ref{tab:transitions}.  \label{fig:mass_hist}}
\end{center}
\end{figure}

%%%%%%%%%%%%%%%%%%%%%%%%%%%%%%%%%%%%%%%%%%%%%%%%
\begin{footnotesize}
\begin{table*}
\begin{center}
 \footnotesize
\centering
%    \hspace*{-1.85 cm}
\begin{minipage}{160mm}
   \caption{Observation data sets of CGM}
   \begin{tabular}{@{}lccccc@{}}
     \hline
     \hline
     \multicolumn{1}{c}{}& \# of Galaxies & $\langle z \rangle^a$  & Impact Parameter Range (kpc)  & $\log{M_* / M_\odot}^b$ & \multicolumn{1}{c}{Transitions} \\
     \hline
     Johnson et al.\ 2015 		& 71 				& 0.271 $\pm$ 0.088 	&	$63-991$  	& 10.14 $\pm$ 0.92 & Ly$\alpha$, O\,VI  \\      
     \hline
     Liang \& Chen 2014 		& 195 			& 0.041 $\pm$ 0.044 	&	$32-499$  	& 9.92  $\pm$ 1.20  & Ly$\alpha$, C\,II, C\,IV, Si\,II, Si\,III, Si\,IV  \\ 
     \hline
     Bordoloi et al.\ 2014 		& 43 				& 0.027 $\pm$ 0.023 	&	$14-135$  	& 9.5 $\pm$ 0.48 & C\,IV  \\      
     \hline
     Werk et al.\ 2013 			& 44 				& 0.221 $\pm$ 0.053 	&	$18-154$  	& 10.61 $\pm$ 0.50 & Ly$\alpha$, C\,II, Si\,II, Si\,III, Si\,IV \\   
     Tumlinson et al.\ 2011		& 				&					&				&				& O\,VI \\
     \hline
     Chen et al.\ 2010 			& 75 				& 0.239 $\pm$ 0.094 	&	$8.5-119$  	& 9.9 $\pm$ 0.58 & Mg\,II  \\      
     \hline
     Steidel et al.\ 2010  		& 512 (stacked)$^c$ 	& 2.2     $\pm$ 0.3 		&	$10-125(280)^d$ & 9.85 $\pm$ 0.46$^e$ & Ly$\alpha$, C\,II, C\,IV, Si\,II, Si\,IV  \\    
     \hline    
     \multicolumn{6}{l}{$^a$Median redshift and dispersion of the galaxy samples.}\\
     \multicolumn{6}{l}{$^b$Median stellar mass and dispersion of the galaxy samples.}\\
     \multicolumn{6}{l}{$^c$Mean equivalent width $W_r$ measured from stacked background galaxy spectra.}\\
    \multicolumn{6}{l}{$^d$Maximum impact parameter is 280 kpc for Ly$\alpha$ and 125 kpc otherwise. }\\
     \multicolumn{6}{l}{$^e$Median and dispersion of stellar mass for \protect\cite{Steidel2010}, a representative sample of galaxies in \protect\cite{Reddy2012}.}\\
 \label{tab:transitions}
    
 \end{tabular}
\end{minipage}
\end{center}
\end{table*}
\end{footnotesize}
%%%%%%%%%%%%%%%%%%%%%%%%%%%%%%%%%%%%%%%%%%%%%%%%

To compare our simulations with observations, we have compiled a set of existing measurements of the CGM absorbers probed with different ions around galaxies with stellar mass estimates. The data set is summarized in Table \ref{tab:transitions}, which specifies the number of galaxies, their mean redshifts and stellar masses, and the range of impact parameters probed by a particular sample. 
The absorber galaxy samples span
a wide range of stellar masses from dwarf galaxies with $M_* \approx 10^7 M_{\odot}$ to $L_*$ galaxies with $M_* \approx 10^{11} M_{\odot}$ at $z \approx 0$, and Lyman-break galaxies with $\langle M_* \rangle \approx 10^{11} M_{\odot}$ at $z \approx 2$. 
Furthermore, Figure \ref{fig:mass_hist} shows the ranges of virial halo masses for the galaxies which generated a particular ion absorber, where virial masses were estimated from galaxy stellar masses using our adopted stellar mass--halo mass relation (see below). Figure~\ref{fig:mass_hist} shows that although the overall range of halo masses is broad, individual ions may be probed by a narrower range of halo masses.

In this section, we first discuss details of how the comparison between observations and simulations is made. We then compare model predictions with observed column density, equivalent width and covering fraction profiles around galaxies and discuss the implications for galaxy formation and feedback.  

\begin{figure*}
\begin{center}
\includegraphics[scale=0.56]{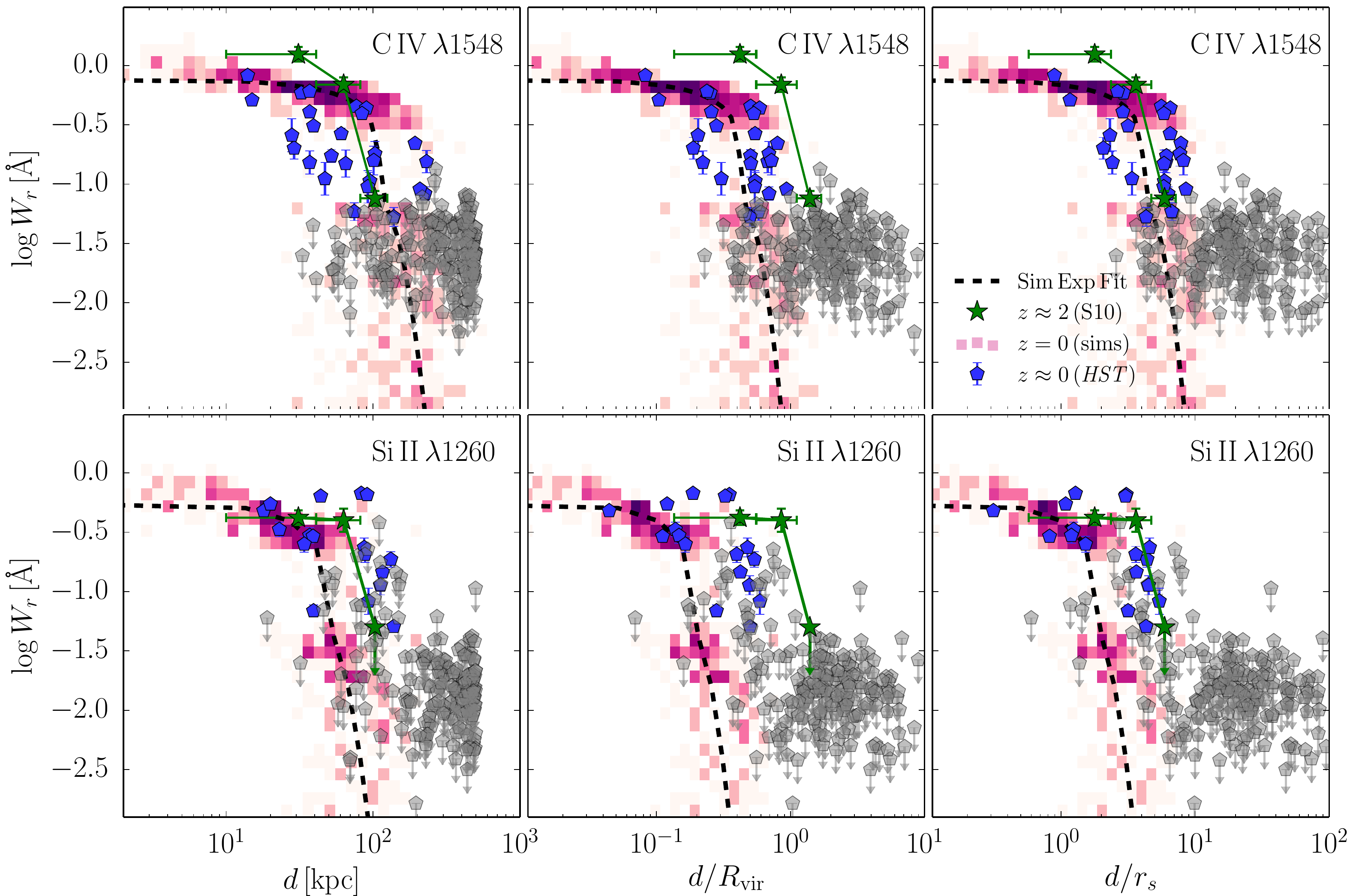}
\caption{Equivalent width distribution with impact parameter for the low-$z$ galaxies observed by the COS instrument onboard \textit{HST} (see Table \ref{tab:transitions} for description of observational samples and references) and $z\approx 2$ stacked measurements by \protect\citet{Steidel2010}.  Blue pentagons are detections, while gray pentagons with arrows are upper limits. The color 2D histogram shows distribution of equivalent widths for the lines of sight through the halo of the MW-sized galaxy in the {\tt ALL\_Efb\_e001\_5ESN} run, assuming turbulent broadening of $b_{\rm NT} = 20 \, \kms$.  Black dashed lines represent $W_r$ profiles converted from the best fit exponential profiles to the simulated column density profiles for corresponding ions. The left panel shows profiles against the impact parameter $d$ in physical kpc, while in the middle and right panels $d$ is rescaled by the virial radius, $R_{\rm vir}$, and halo scale radius, $r_s$, respectively. \label{fig:Wrevol}}
\end{center}
\end{figure*}

\begin{figure*}
\begin{center}
\includegraphics[scale=0.56]{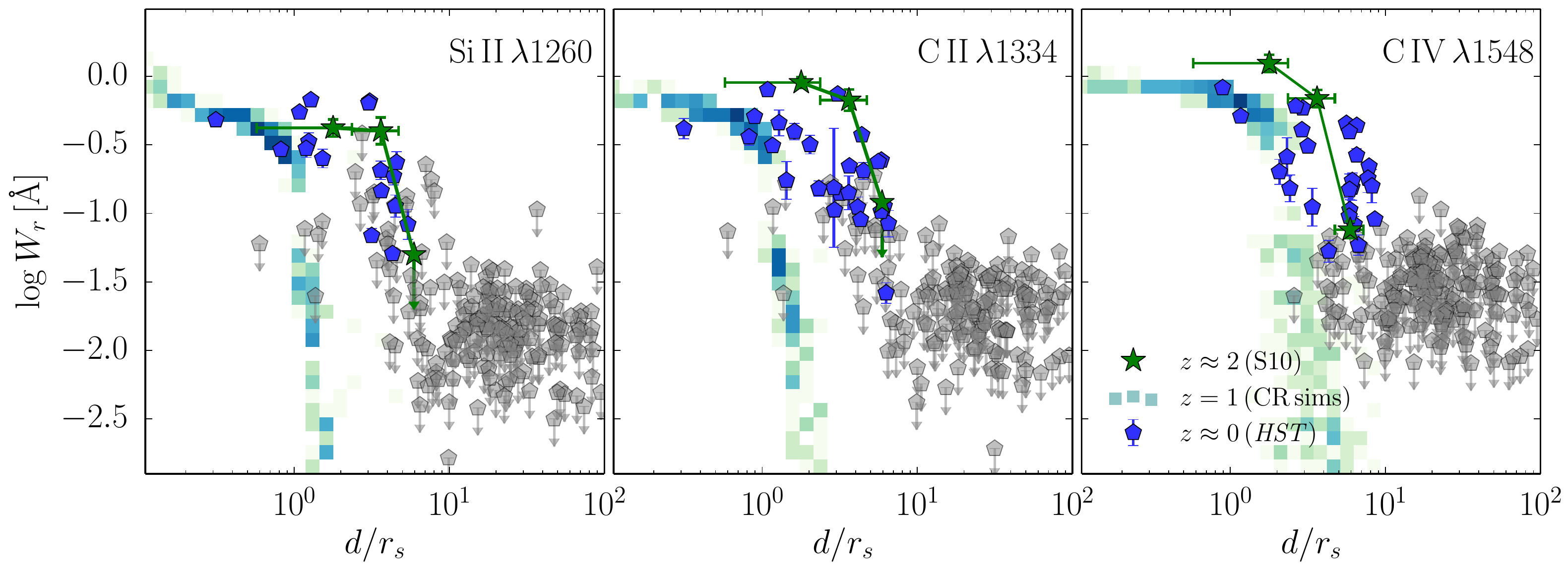}
\caption{Equivalent width profiles for three representative ions in the run with cosmic ray feedback at $z = 1$ as a function of impact parameter scaled by the scale radius, $d/r_s$ (2D color histogram) compared to observations (points; the point types are the same as in Figure \ref{fig:Wrevol}). \label{fig:Wrevol_CR}}
\end{center}
\end{figure*}

\subsection{Rescaling observations and simulations for fair comparison} 
\label{sec:rescalingsimobs}

Observations of  low-$z$ absorbers in our data set span more than four decades of stellar mass  \citep{Chen2010, Tumlinson2011, LiangChen2014, Werk2014,Bordoloi2014, Johnson2015}, while our simulations model a single progenitor of mass close to $L_*$ galaxies. We thus need to rescale both data and simulations in a way that makes the comparison sensible. 

In the previous section, we showed that the column density profiles in simulations scale most optimally across redshifts not with the virial radius of their parent halo, but with the scale radius of their halo density profile, defined as the radius where the profile has logarithmic slope of $-2$. To test whether a similar scaling applies to observed profiles in Figure \ref{fig:Wrevol} we show the equivalent width profiles of absorbers in the {\tt ALL\_Efb\_e001\_5ESN} simulations along with observational data at redshifts $z\in 0-0.3$ and $z\approx 2$ with radii in physical kpc, and rescaled by the virial radius of host halo $R_{\rm vir}(z)$ and $r_s(z)$. We show evolution of the $W_r$ profiles for $\rm C\,IV$ and $\rm Si\,II$ only, because the profiles for these ions exhibit some of the sharpest ``edges'' and because results for other ions are similar. 

We convert our predicted column density into rest frame equivalent width $W_{\rm{r}}$ using the curve of growth analysis discussed in section 3.3. To account for turbulence not fully resolved
in simulations in the relatively low resolution regions of gaseous halos, we use the $b$ parameter as $b^2 = b_{\rm{T}}^2 + b_{\rm{NT}}^2$, and adopt $b_{\rm NT} \approx 20 \kms $ \citep[see also][]{Oppenheimer2012}, although we checked that using $b_{\rm NT}$ also provides a reasonable match to observations.    

In the simulations the virial radius of the galaxy is known from the density profile of the matter distribution. In observations, however, we need to estimate it statistically using stellar masses. We do this using the average stellar mass-halo mass relation derived by \citet{Kravtsov2014} using the abundance matching. This relation uses the measurement of galaxy stellar mass function by  \citet{Bernardi2013}, which is based on updated photometry of the SDSS main galaxy sample that corrects for background subtraction errors in the previous SDSS data releases. 

We assume a non-evolving $M_*-M_{\rm vir}(z)$ relation \citep[e.g.,][]{Behroozi2013}
 to convert $M_*$  to $M_{\rm vir}$ for each observed 
galaxy in the dataset. We then use $M_{\rm vir}$ to compute virial radius using its definition $R_{\rm vir}(z)=[3M_{\rm vir}(z)/(4\pi\Delta_{\rm vir}\bar{\rho}(z))]^{1/3}$, where 
$\Delta_{\rm vir}(z)$ is the redshift-dependent ``virial'' density contrast computed using approximation to the spherical collapse overdensity given by  \cite{BryanNorman1998}.\footnote{Adopting other commonly used definitions of the ``virial radius'', such as $R_{\rm 200m}$ or $R_{\rm 200c}$, would not qualitatively affect our conclusions.}
The scale radius is obtained from the virial radius assuming the median concentration of halos of mass $M_{\rm vir}$ at redshift $z$, predicted by the model of \cite{DiemerKravtsov2015}:
$r_s=R_{\rm vir}/c_{\rm vir}(M_{\rm vir},z)$.

Figure \ref{fig:Wrevol} shows that evolution in the physical radius from $z\approx 2$ to the present epoch in the observational sample is quite mild \citep[as found previously by][]{Chen2012,LiangChen2014}.  The figure also shows that the drop in the $W_r$ profiles is sharper when $r_s$ rescaling is used instead of $R_{\rm vir}$. 
Furthermore, the profile rescaled by $r_s$ does not evolve significantly between $z\approx 0$ and $z\approx 2$. 

This alignment of the profiles at different $z$ is remarkable because properties  of the low-$z$ and $z\approx 2$ populations of galaxies are drastically different. For example, typical star formation rates differ by over an order of magnitude.  Given the theoretical arguments for $r_s$ scaling and the fact that it appears to work for real CGM, we will scale impact parameters in the simulations and observational sample
by the  radius of $4r_s$ in the comparisons presented below. We choose the multiple of four because at $z\approx 2$ this radius is actually close to the virial radius of galaxy halos \citep[see][for a detailed discussion]{more2015}.

\begin{figure*}
\begin{center}
\includegraphics[scale=0.8]{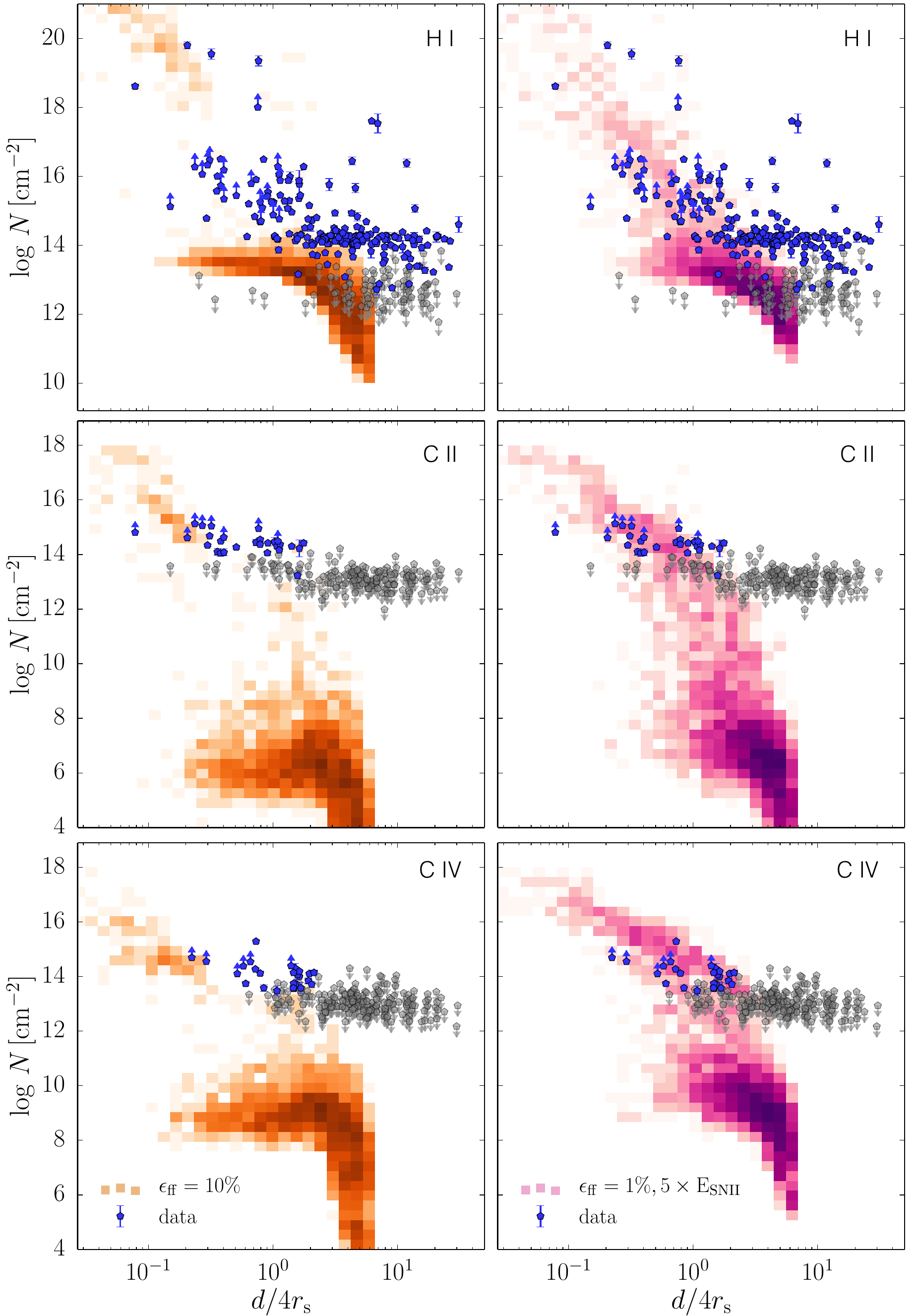}
\caption{Comparison between predicted radial column density distributions of various ion absorbers (2D histograms) with observations at $z \approx 0 - 0.3$ (pentagon points; see Table \ref{tab:transitions}). Blue points show detections, while gray points show non-detections with downward arrows denoting $2\sigma$ upper limits. The ion for which the profile is constructed is indicated in the upper right. The left column shows profiles predicted by the fiducial model with all feedback $+$ $E_{\rm{fb}}, \epsilon_{\rm{ff}} = 10\%$. The right column shows the predicted profiles from the model with all feedback $+$ $E_{\rm{fb}}, \epsilon_{\rm{ff}} = 1\%, 5E_{\rm{SNII}}$.  The 2D histograms show that the majority of the points in \texttt{ALL\_Efb\_e010} are many orders of magnitude below observations, while profiles in the \texttt{ALL\_Efb\_e001\_5ESN} run are closer to observations.  \label{fig:NdRone}}
\end{center}
\end{figure*}

\begin{figure*}
\begin{center}
\includegraphics[scale=0.8]{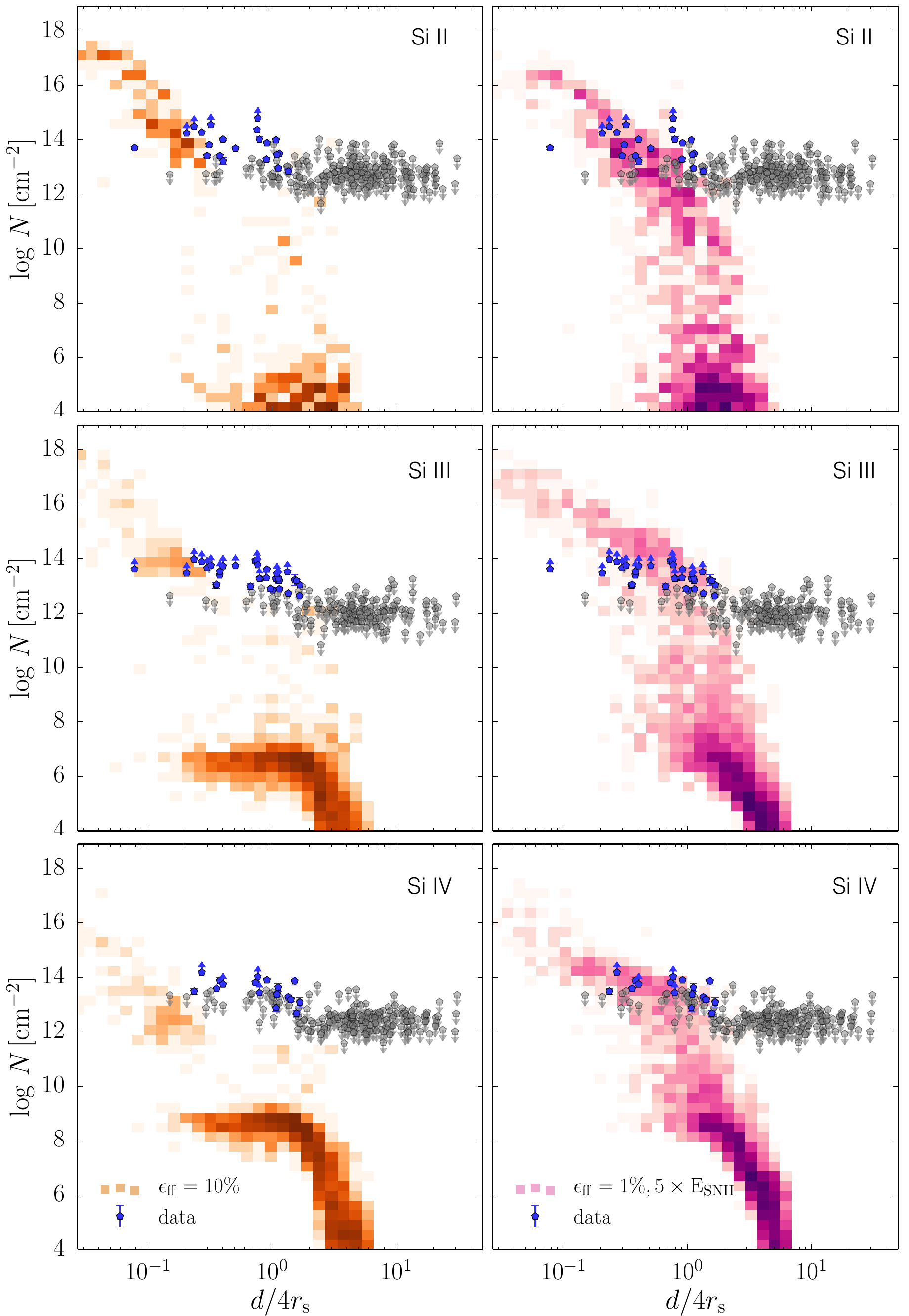}
\caption{Similar to Figure \ref{fig:NdRone}, but for Si\,II, Si\,III and Si\,IV.   \label{fig:NdRtwo}}
\end{center}
\end{figure*}

\begin{figure*}
\begin{center}
\includegraphics[scale=0.8]{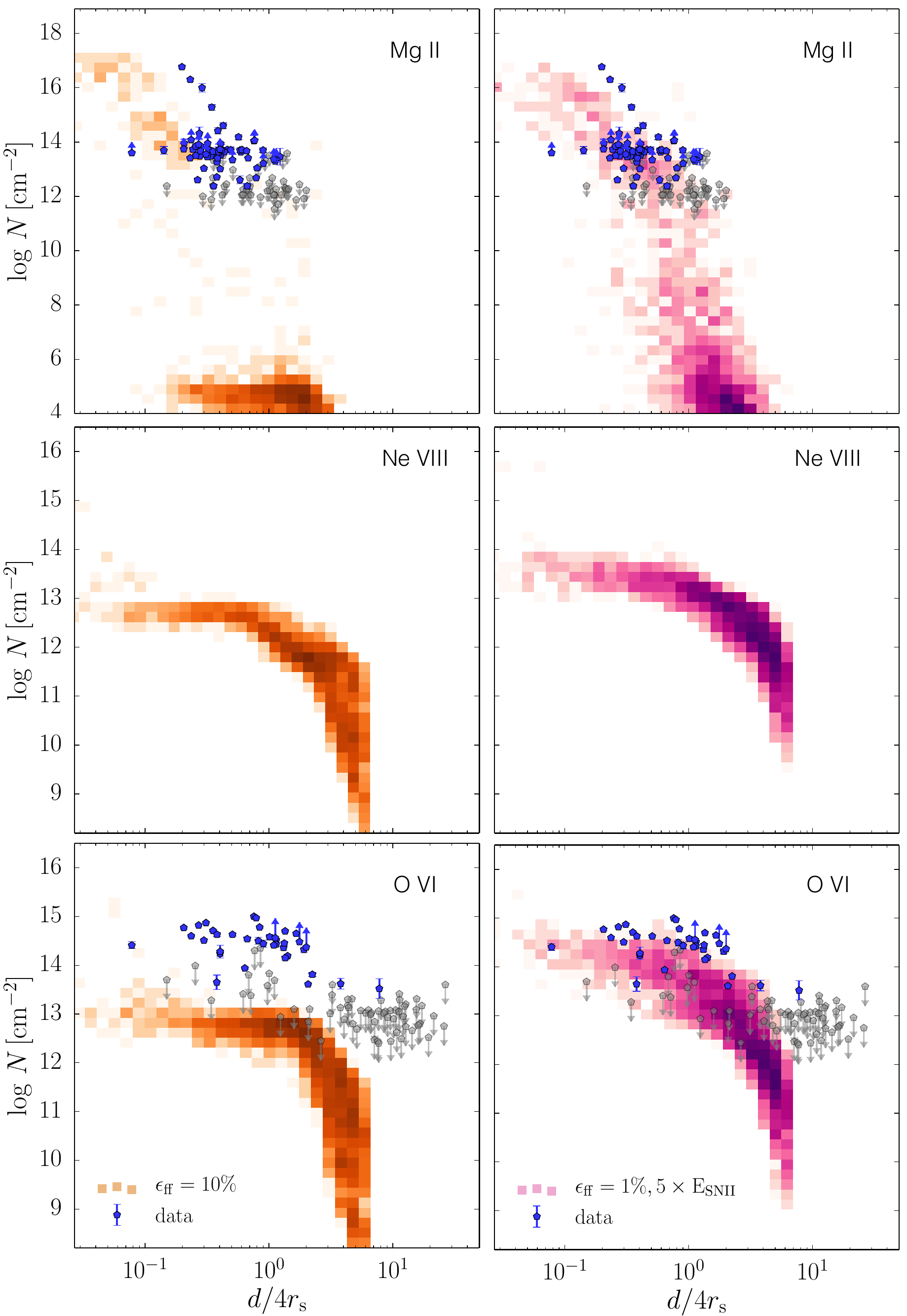}
\caption{Similar to Figure \ref{fig:NdRone}, but for Mg\,II, O\,VI and Ne\,VIII.  \label{fig:NdRthree}}
\end{center}
\end{figure*}

\subsection{Comparison of the column density profiles in simulations and observations}

Figures \ref {fig:NdRone}, \ref {fig:NdRtwo} and \ref {fig:NdRthree} compare column density distributions with $d/4r_s$ in the two simulations that were run to $z=0$,  \texttt{ALL\_Efb\_e010} and {\tt{ALL\_Efb\_e001\_5ESN}},  and observations for all commonly observed transitions.  It is immediately apparent that the absorber column densities in  the fiducial run \texttt{ALL\_Efb\_e010} with $\epsilon_{\rm{ff}} = 10\%$ for all ions are consistently low by up to several orders of magnitude compared to observations. Interestingly, \citet[][]{AgertzKravtsov2015, AgertzKravtsov2016} show that this run  predicts properties of the stellar and cold HI gas component of the galaxy in very good agreement with properties of a typical late type galaxy of this mass. Thus, there is a striking difference in how successful this run is 
in producing realistic stellar components of galaxies and the discrepant gaseous CGM it predicts. 

In contrast,  the \texttt{ALL\_Efb\_e001\_5ESN} run matches the CGM column density distribution of all ion species in the $\log N - d/4r_s$ plane reasonably well, especially given that we are comparing a single object to galaxies of a range of stellar masses. However, this run produced stellar
component of the galaxy with a spheroidal morphology, exponential stellar surface density profile and unrealistically large half-mass radius.  \textit{This shows that the CGM provides a stringent and orthogonal constraint on the galaxy formation simulations relative to the optical observations of galaxies}.

It is interesting that  the CGM in the \texttt{ALL\_Efb\_e001\_5ESN} run reproduces the 
sharp drop in the column density of absorbers apparent  for all transitions, which coincides with the location where a large fraction of measurements are upper limits.  \citet{LiangChen2014} have estimated this drop to be at $d/R_{\rm{vir}} \approx 0.7$ for Si\,II, Si\,III Si\,IV, C\,II and C\,IV. A similar drop of  the C\,IV column densities was reported by \citet{Bordoloi2014} at $d/R_{\rm{200m}} \approx 0.5$, and for O\,VI by \cite{Johnson2015} at $d/R_{\rm{vir}} \approx 1$. 

The difference between $d/R_{\rm{200m}} = 0.5$ and  $d/R_{\rm{vir}} = 0.7$ found for low and intermediate ions is simply due to the difference between $R_{\rm vir}$ ($\Delta\approx 360$) and $R_{\rm 200m}$ definition  and the specific adopted $M_* - M_h$ relation. Once consistent definitions are adopted, the two samples of data are in agreement. Indeed, if we re-compute $R_{\rm{vir}}$ using the Bryan \& Norman approximation and \citet{Kravtsov2014} $M_* - M_h$ for all galaxies compiled in this study, the low-ionization metal boundary is consistent with $d/R_{\rm{vir}} \approx 0.7$. 

\citet{LiangChen2014} found a hint that the  C\,IV extends beyond $d/R_{\rm{vir}}  = 0.7$, but based only on
absorbers around two galaxies. Here, combining the data sets of \citet{LiangChen2014} and \citet{Bordoloi2014}, we see a stronger indication that C\,IV indeed  extends to $d\approx 6-7 r_s \approx 0.9 R_{\rm vir}$ and its distribution is thus more extended compared to the low-ions (e.g., Si\,II and Mg\,II). Taking into account the result of  \citet{Johnson2015} that the O\,VI distribution extends to $d\approx R_{\rm{vir}}$, observations indicate that the extent of the absorbing gas increases with increasing ionization energy of the ions.  This is qualitatively consistent with the tight trend between the scale height of the exponential column density profile and ionization energy of the corresponding ion that we found in simulations (see Figure \ref{fig:hEion} and Table \ref{tab:Eion}).  It remains to be seen whether observations will confirm the tight correlation of the profile scale height with ionization energy predicted by our simulations. 

\subsection{Comparing the CGM Covering Fractions}

The scatter of the column density in the region of sharp drop of ion column densities shown in Figures \ref{fig:NdRone}, \ref{fig:NdRtwo} and \ref{fig:NdRthree} is large, and at some radii may span orders of magnitude. The large scatter means that it may be misleading to compare median or average profiles and one has to compare distributions of column densities directly, as we did above. Another way to characterize the column density distribution is via the covering fraction $\kappa$, which measures a fraction of area covered by absorbers with column density or equivalent width larger than a given threshold. 

Different detection thresholds reported for different observations present a challenge for a uniform comparison. Some studies adopt rest frame equivalent width of $W_0 = 0.05 - 0.1 $ \AA\, and $\log N_0 = 13.5\ \rm{cm^{-2}}$ as detection thresholds \citep{LiangChen2014,Johnson2015}. We adopt these values here for comparison with the derived covering fraction of data from \citet{LiangChen2014} and \citet{Johnson2015} for O\,VI. For observations where only equivalent width measurements are available, \citet{LiangChen2014} adopted the threshold $W_r = 0.05 $ \AA. We convert this value from $W_0$ to $\log N_0 = 13.75$ for the calculation of the covering fraction, using their estimated $b$ parameter of 29 \kms. 

\begin{figure*}
\centering
\includegraphics[scale=0.55]{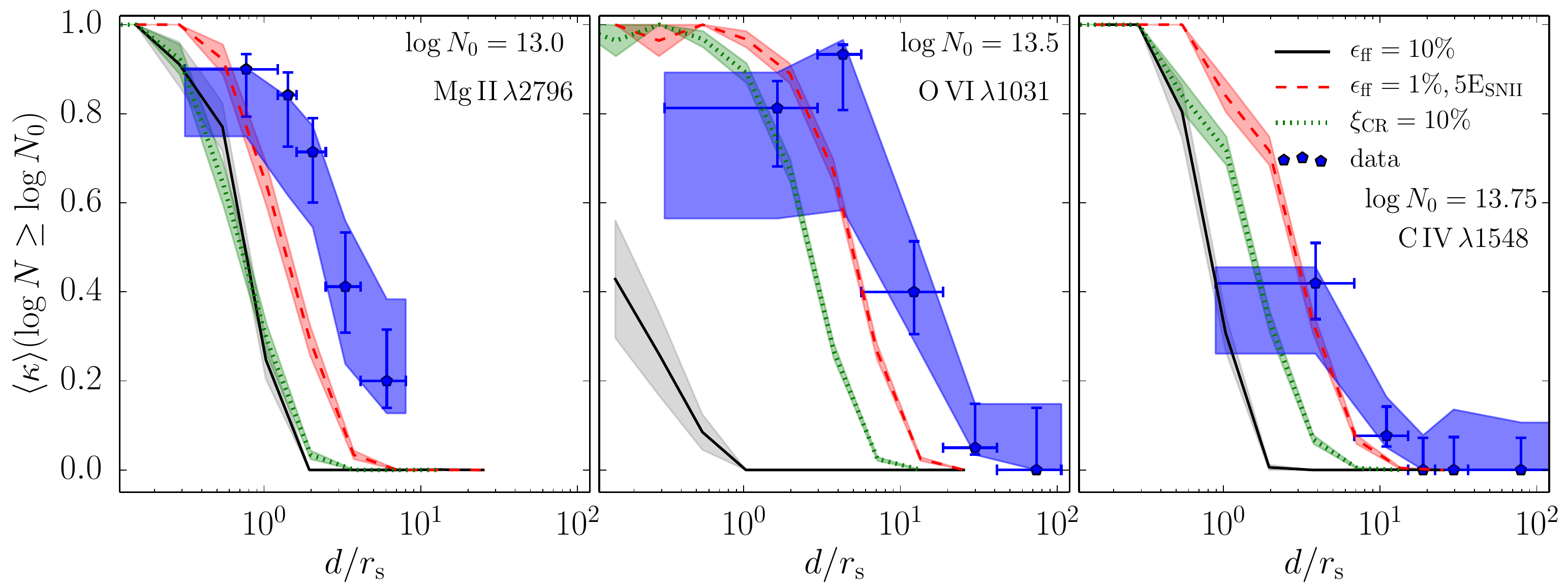}
\caption{Comparison of the predicted covering fraction profiles with measurements of Mg\,II $\lambda 2796$ \protect\citep{Chen2010}, C\,IV $\lambda 1548$ \protect\citep{LiangChen2014, Bordoloi2014} and O\,VI $\lambda 1031$ \protect\citep{Tumlinson2011, Johnson2015} transitions.  Detection threshold $W_0 (\rm{C\,IV} \lambda 1548) = 0.05$ \AA\, is adopted from \protect\cite{LiangChen2014}. We convert the threshold to column density using a curve-of-growth analysis adopting the doppler parameter $b = \sqrt{2} \sigma = 29 \kms $. Detection threshold $\log N_0 (\rm{O\,VI}\,\lambda 1031)= 13.5 \, \rm{cm^{-2}} $ is adopted from \protect\cite{Johnson2015}. $\log N_0 (\rm{Mg\,II} \lambda 2796) = 13 \, \rm{cm^{-2}}$ is chosen based on the level upper limits in \protect\cite{Chen2010}. Error band represents  68\% confidence level bracketed by two limiting cases of consideration of upper limits above the detection threshold (e.g., as detection and as non-detection).  In all three ions, \texttt{ALL\_Efb\_e001\_5ESN} and \texttt{ALL\_e010\_CR} predictions agree with the inner region ($< 2 - 4 r_{\rm s}$) of the data while under-predicting the covering fraction of the outskirts of the halo ($\sim 10 r_{\rm s}$). Note that \texttt{ALL\_e010\_CR} is at $z = 1$ while the other runs are at $z = 0$. It is conceivable that halo in the CR run will continue to grow at lower redshift, as suggested by Fig. \ref{fig:MgIIpro} \label{fig:coverfrac} }
\end{figure*}

In principle, one can use a different way of treating the upper limits in which the upper limit is considered either as detection with the corresponding column density or equivalent width, or as effectively zero column density. The former choice gives the largest possible covering 
fraction, while the latter gives the smallest possible covering fraction. These two limits are shown as blue bands in the panels of Figure \ref{fig:coverfrac}. Overall, the two methods give estimates consistent with each other, but the second approach in some cases results in a wider range of allowed covering fractions. 
We also perform a bootstrap re-sampling of the simulations to estimate the uncertainty of the covering fraction profile in simulated galaxies. 

Figure \ref{fig:coverfrac}  shows the covering fraction profile predictions for Mg\,II,  C\,IV and O\,VI for runs {\tt ALL\_Efb\_e010}, {\tt ALL\_e010\_CR} and {\tt ALL\_Efb\_e001\_5ESN}. As discussed above, the column densities of most lines of sights in the {\tt ALL\_Efb\_e010} run is orders of magnitude too low compared to observations. These low column densities are below the detection limit of the observations, resulting in low covering fraction.
This is apparent for Mg\,II, C\,IV and even O\,VI. The only region with non-zero $\kappa$ is with distance $d < r_s$ (or $\approx 0.1 R_{\rm vir}$ for the simulated halos),  at the extended disk where density is high and temperature is low. 

On the other hand, the {\tt ALL\_Efb\_e001\_5ESN} run produces cold/warm outflows that reach to larger distances, producing a higher covering fraction.  The covering fraction profile predicted in 
this run is much closer to the profile derived from observations. Although the match is not perfect, we note that we are comparing a single object in simulations to a sample of galaxies spanning
a wide range of stellar masses. Thus, meaningful conclusions can be drawn only from larger samples of simulated galaxies. Nevertheless, the fact that simulations reproduce high covering 
fraction extent to large radii in rough agreement with observations is encouraging. 

The run {\tt ALL\_e010\_CR}, which includes cosmic ray feedback, predicts profiles at $z=1$ (the lowest $z$ to which the simulation was run) that are intermediate between those of {\tt ALL\_Efb\_e010} and {\tt ALL\_Efb\_e001\_5ESN}. For O\,VI it is considerably closer to {\tt ALL\_Efb\_e001\_5ESN}. This indicates that cosmic ray-aided winds can help to bring predictions in better agreement with observations.

%-----------------------
\section{Discussion}
\label{sec:discussion}
%-----------------------

In the preceding section we have presented comparisons of properties of gaseous halos forming around a progenitor of a $\approx L_*$ 
galaxy in simulations with different parameters and implementations of the star formation -- feedback loop.  
We find that properties of the gaseous CGM, in particular, spatial distribution of various ions, are highly sensitive to the 
details of these implementations and parameter choices.  For example, the fiducial simulation of \citet[][]{AgertzKravtsov2015,AgertzKravtsov2016}, which produces a very realistic central galaxy with the correct stellar mass, size, angular momentum, rotation curve, bulge-to-disk ratio, stellar and gas surface density profiles does not produce extended CGM and is in striking discrepancy with observations. This illustrates that properties of galaxies and properties of their CGM provide strong {\it complementary} constraints on the processes governing galaxy formation. 

Variations of the stellar feedback model, such as adding feedback modelling due to cosmic rays or simply changing parameters governing star formation and stellar feedback, affect properties of gaseous halos around simulated galaxies and produce a more extended CGM in better agreement with observations (see, e.g., Figs. \ref{fig:NdRone}-\ref{fig:NdRthree} above). This is particularly true for ions with higher ionization energies, such as C\,VI and O\,VI, indicating that these simulations correctly capture the thermal and density structure and metallicity profile of hotter gas. 

There are indications that the distribution of low-ions in all of our simulations is somewhat less extended than indicated by observations. In the fiducial run, this is due to the lack of low temperature gas at large radii at low $z$ \citep[see also][]{Hummels2013}. This is mainly because in the fiducial run there are no global winds at $z<0.5-1$ and thus there is no cold/warm gas lifted from the disk of the galaxy into the halo. In contrast, in the 5$E_{\rm SN}$ run, the winds continue to $z = 0$ and therefore the CGM is filled with cold/warm gas at low redshifts.

The low metallicity of the halo in that run may also play a role in the low column density of absorbers. For example, the total column density of all carbon 
 is still lower than the observed column density of CIV alone by a factor of $\sim$10. We think this is because the winds at high $z$ in the fiducial run are so strong that they disperse metals over a large volume, lowering the overall metallicity.

Some recent observations indicate the presence of low-ion absorbing clouds with sizes below the resolution of the simulations \citep{Muzahid2014,Crighton2015}, which is particularly poor in the gaseous halo far away from the dense regions of the disk. It is possible that a sub-grid model of small clouds created by nonlinear thermal instabilities \citep{Joung2012} can ease the disagreement. Gas inflow along filaments and feedback-driven outflows could plausibly excite such instabilities, which then could create clouds that would boost the amount of cold/warm gas present. Some low-redshift observations, on the other hand, indicate large physical sizes \citep{Davis2015} and non-hydrostatic states of many absorbers \citep{Werk2014} consistent with the properties of recent outflows of the kind we observe in our \texttt{ALL\_Efb\_e001\_5ESN} and \texttt{ALL\_e010\_CR} runs. Low-ion absorbers in galaxies thus may originate both from large-scale outflows and from small clouds forming by thermal instability that outflows help to excite. 

Overall, the comparison of simulation results with different star formation and feedback models has produced valuable insights about the radial distribution of absorbers, which we discuss below. Below we will also discuss how the results of our simulations compare to other recent theoretical studies on this subject. 

%------------------------------------------------------------------------------------------------------------------
\subsection{Column density profiles of the CGM absorbers: is there an ion boundary of the CGM?}
\label{sec:expboundary}
%------------------------------------------------------------------------------------------------------------------

As we discussed in section \ref{sec:rescalingsimobs}, observations of absorption lines of a variety of ions, such  as Si\,II, Si\,III Si\,IV, C\,II, C\,IV, and O\,VI, exhibit a sharp drop in the incidence of absorber detections and their covering fraction beyond a certain radius \citep{LiangChen2014, Bordoloi2014,Johnson2015}.  Some researchers have interpreted this drop-off as an ``ion boundary.'' 

 Our simulations  of a $\sim L_*$ galaxy progenitor at different redshifts predict approximately exponential column density profiles  for all ions. When plotted on a log-log scale, the exponential profiles exhibit a sharp turnover at approximately the ``half-mass radius" or $\approx 1.68h_s$, where $h_s$ is the scale height of the exponential profile. Thus, our simulations indicate that the perceived sharp turnover in the incidence
of high column density absorbers is simply a manifestation of a continuous underlying exponential profile. 

It is worth noting that de-projection of 
the projected exponential column density profile via the Abel integral gives the modified Bessel function of the second kind, $K_0(x)$ for the radial profile of 3D number density \citep[see, e.g., Appendix in][]{PittsTayler1997}. The $K_0(r)$ profile has a shallow slope at small $r$, which steepens at larger $r$ and we find that the $K_0(x)$ function does approximate the gas density profile in our simulations at $r\sim 20-200$ kpc
quite well. The origin of the approximately exponential total column density profile, $N_{\rm H}(d)$, is thus in the shape of the 3D gas density profile. 

The column density profiles of specific ions are additionally shaped by the metallicity profile and by the ionization factor and thermal structure of the gas that affect their photo- and collisional ionization rates. These
factors are ion-specific and give rise to profiles that vary systematically with the ionization energy of ions.
 The metallicity profile in the central regions varies in different simulations. In the \texttt{ALL\_Efb\_e001\_5ESN} run the metallicity rapidly increases with decreasing radius at $r\lesssim 50$ kpc, while in the fiducial
 {\tt ALL\_Efb\_e010}  run the metallicity remains low and its profile remains flat down to small radii. The difference is that the former simulation drives outflows at low redshifts that enrich the central region ($\sim 50$ kpc) and provides a steady supply of fresh warm gas that gives rise to low-ion absorbers. 
 
 \begin{figure}
\centering
\includegraphics[scale=0.5]{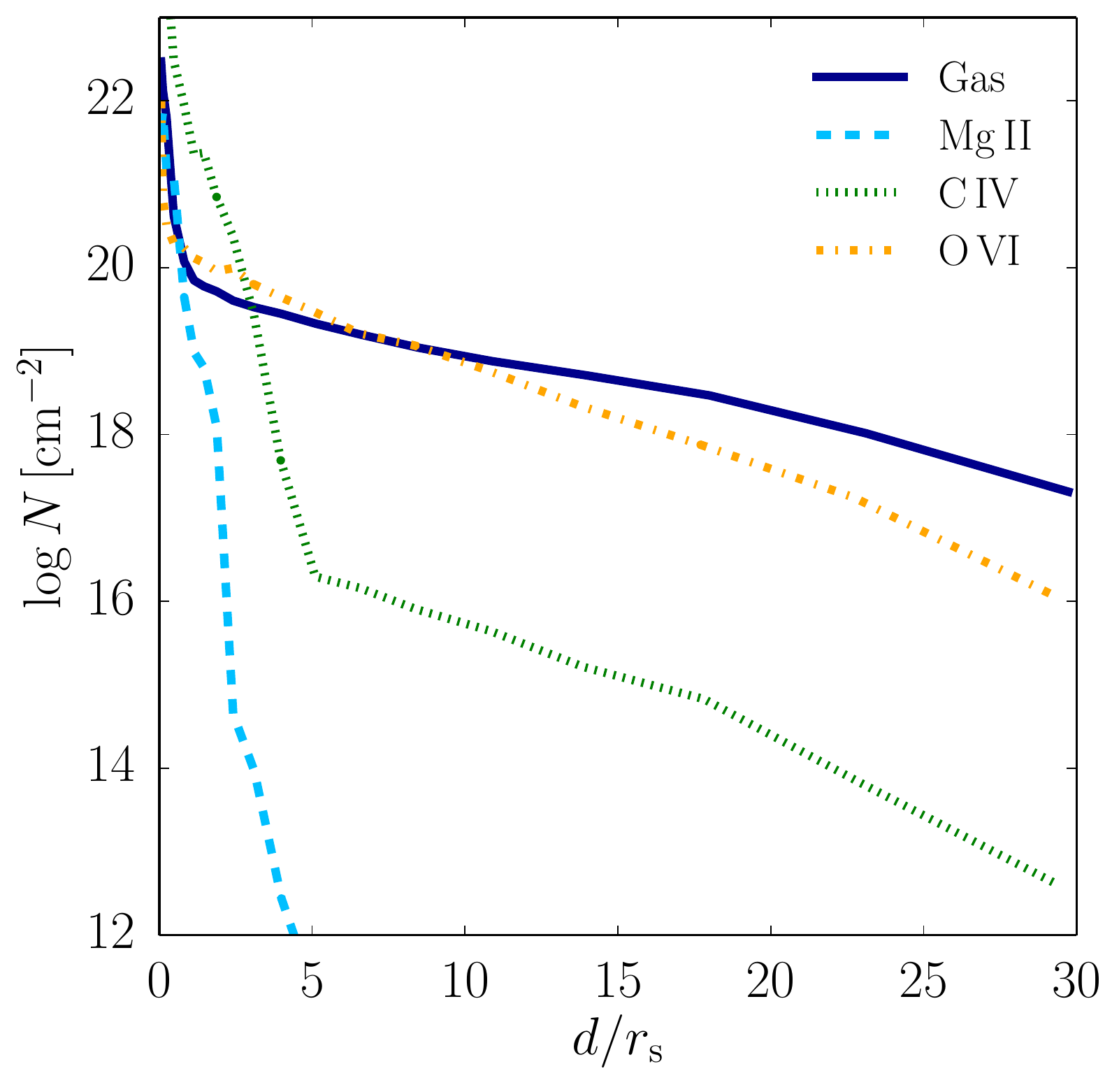}
\caption{Column density profiles of Mg\,II, C\,IV and O\,VI in the \textit{outer region} of the hot gaseous halo share the same underlying exponential profile traced by the total gas column density profile. All other ions (not shown) also show similar exponential profiles. Column densities for ions have been offset by  $\log N_0 = 6$.\label{fig:Nprofiles} }
\end{figure}

We find that column density profiles of all ions trace the overall shape of $N_{\rm H}(d)$ quite closely at $d\gtrsim 50$ kpc, as shown in Figure~\ref{fig:Nprofiles}, because the metallicity profile at these radii in our simulations is quite flat.
The low-ions at the large radii have column densities much below the sensitivity limits of observations. However, higher ionization ions, such as O\,VI, do produce detectable absorption.  Thus, the exponential column density profile of ions like O\,VI simply reflects the underlying total gas profile. At $r\lesssim 2r_s\approx 50$ kpc the profiles are more difficult to interpret, but it appears that the cycling outflows at these radii do result in a metallicity profile 
 that is also approximately exponential in projected radius. Other factors, such as dependence of the $U$-factor on radius, probably also contribute to the shape of the column density profiles of specific ions. In particular, the U-factor almost centrainly plays a role in setting the tight correlation of scale height with ionization energy we find for the low ions.  

Encouragingly, there have been a few recent studies that show exponential profiles are a good fit to the observed data. Study by \citet{Nielsen2013} finds that an exponential profile provides a better fit to the Mg\,II equivalent width profile, as compared to a power law. \citet{Borthakur2015} finds an approximately exponential form for  the Ly$\alpha$ equivalent width profile. \citet{Bordoloi2014} have also used an exponential function to describe the sharp drop in the equivalent width of CIV absorber column density in their sample, although overall they used a more complicated profile given by a product of a power law and an exponential function. 
  
%------------------------------------------------------------------------------------------------------------------
\subsection{Column density profiles of the CGM absorbers: scaling with halo mass and redshift}
\label{sec:proscaling}
%------------------------------------------------------------------------------------------------------------------

Interestingly, as shown in section \ref{sec:simobs} above, the collection of observations considered in our study indicates that the scale height of the column density profiles 
for observed absorbers around galaxies of a wide range of stellar masses occurs at approximately the same multiple of the halo scale radius, $r_s$ at different redshifts. The transition in galaxies at $z\approx 2$ in the sample of \citet{Steidel2010} also occurs at the same multiple of $r_s$. Regardless of the precise form of the profile, this scaling is consistent with the scaling predicted in our simulations for the inner regions of the CGM profiles. 
 
The scaling of the overall column density profile with $r_s$ is due to the slow evolution of the inner gaseous profiles. 
Simulations of halo evolution in CDM models show that the inner profiles of galaxy-sized halos are in equilibrium and
evolve little after $z\approx 1-2$  \citep{prada_06, diemand_vl_2007, Cuesta2008, Diemer2013, Zemp2014,DiemerKravtsov2014, correa:2015}, while the profiles in the outskirts ($r\gtrsim R_{200c}$) continue to accrete mass and evolve \citep[][see also discussion in Section \ref{sec:simproevol} above]{DiemerKravtsov2014,more2015}. 
The virial radius of the halo tracks the evolution of the outer profile and the outer splashback boundary of the halo fairly closely and thus evolves considerably faster than the profile in the inner regions. The evolution of the latter can be better characterized by the halo scale radius \citep{more2015}, which also exhibits slow evolution at low redshifts \citep{Bullock2001}. If the overall gaseous profile is in approximate equilibrium, this can explain the apparent scaling 
of the column density profile evolution with $r_s$. 

Note that {\it at a fixed redshift}, the scalings with $R_{\rm vir}$ and $r_s$ are nearly equivalent because $r_s\equiv R_{\rm vir}/c_{\rm vir}$ and concentration is a very weak function of halo mass: $c_{\rm vir}\propto M^{-s}_{\rm vir}\propto R^{-0.3s}_{\rm vir}$, where $s\approx 0.08-0.1$ at $z=0$ and is shallower at higher redshifts \citep[e.g.,][]{Bullock2001,Neto2007,DiemerKravtsov2015}. Thus scaling with $r_s$ at a fixed $z$ is equivalent to scaling 
with $R_{\rm vir}^{\approx 1-1.25}$. At the same time, concentration exhibits a rather strong evolution with redshift due to fast evolution of $R_{\rm vir}$ and thus redshift evolution of $r_s$ and $R_{\rm vir}$ and corresponding scaling as a function of redshift will be very different. Therefore, our results indicate that for samples spanning a wide range of 
redshift, scale radius $r_s$ should be used to rescale radial scales, such as the impact parameter. 

The remaining question is why the column density profiles shaped by stellar feedback driven outflows in simulated galaxies scale with $r_s$. We believe this can be understood from the following  considerations. The specific energy of the wind when it is ejected is $E=v_{\rm w}^2/2+\phi(r_{\rm w})$, where $v_{\rm w}$ is initial wind velocity at the launch radius, $r_{\rm w}$ and $\phi$ is gravitational potential at $r_{\rm w}$. If the outflowing gas is gravitationally bound ($E<0$), the wind will stop at some radius $r_{\rm out}<R_{\rm vir}$. If we assume that energy losses due to ram pressure force from the tenuous halo gas on the wind can be neglected, then the wind stops at $E\approx \phi(r_{\rm out})$. If $r_{\rm out}$ is sufficiently large we can approximate the potential by the form expected from the NFW profile:
\begin{equation}
\phi(r_{\rm out})\approx -4.63\,V_{\rm max}^2\frac{\ln(1+x_{\rm out})}{x_{\rm out}},
\end{equation}
 where $V_{\rm max}$ is the maximum circular velocity of the halo, $x_{\rm out}\equiv r_{\rm out}/r_s$ and $r_s$ is the scale radius of the NFW profile. Although the potential in the region where stars launch winds is likely not described by the NFW form due to contributions of the baryons in the galaxy, its overall amplitude should still scale as $\phi(r_{\rm w})\propto -V_{\rm max}^2$. At the same time, for both momentum- and energy-driven winds we expect $v_{\rm w}\propto V_{\rm max}$ \citep[][]{Murray2005,Booth2013,Muratov2015}, the scaling that is also indicated by observations \citep[][although see Heckman et al. 2015, submitted]{SchwartzMartin2004,Rupke2005,Schwartz2006}. Therefore, the above equations suggest that
\begin{equation}
\frac{\ln(1+x_{\rm out})}{x_{\rm out}}\approx c_1 +c_2\Psi (r_{\rm out}),
\end{equation}
where $c_1$, $c_2$ are constants and $\Psi(r_{\rm out})$ does not depend explicitly on $V_{\rm max}$. This equation means that $x_{\rm out}$ is independent of $V_{\rm max}$ and hence halo mass and thus $r_{\rm out}\propto r_s$. Although individual fountains may have a range of initial velocities and launch radii, from the above considerations we can see that the characteristic extent of the profile they shape may naturally scale with $r_s$. 

\begin{figure*}
\begin{center}
\includegraphics[scale=0.5]{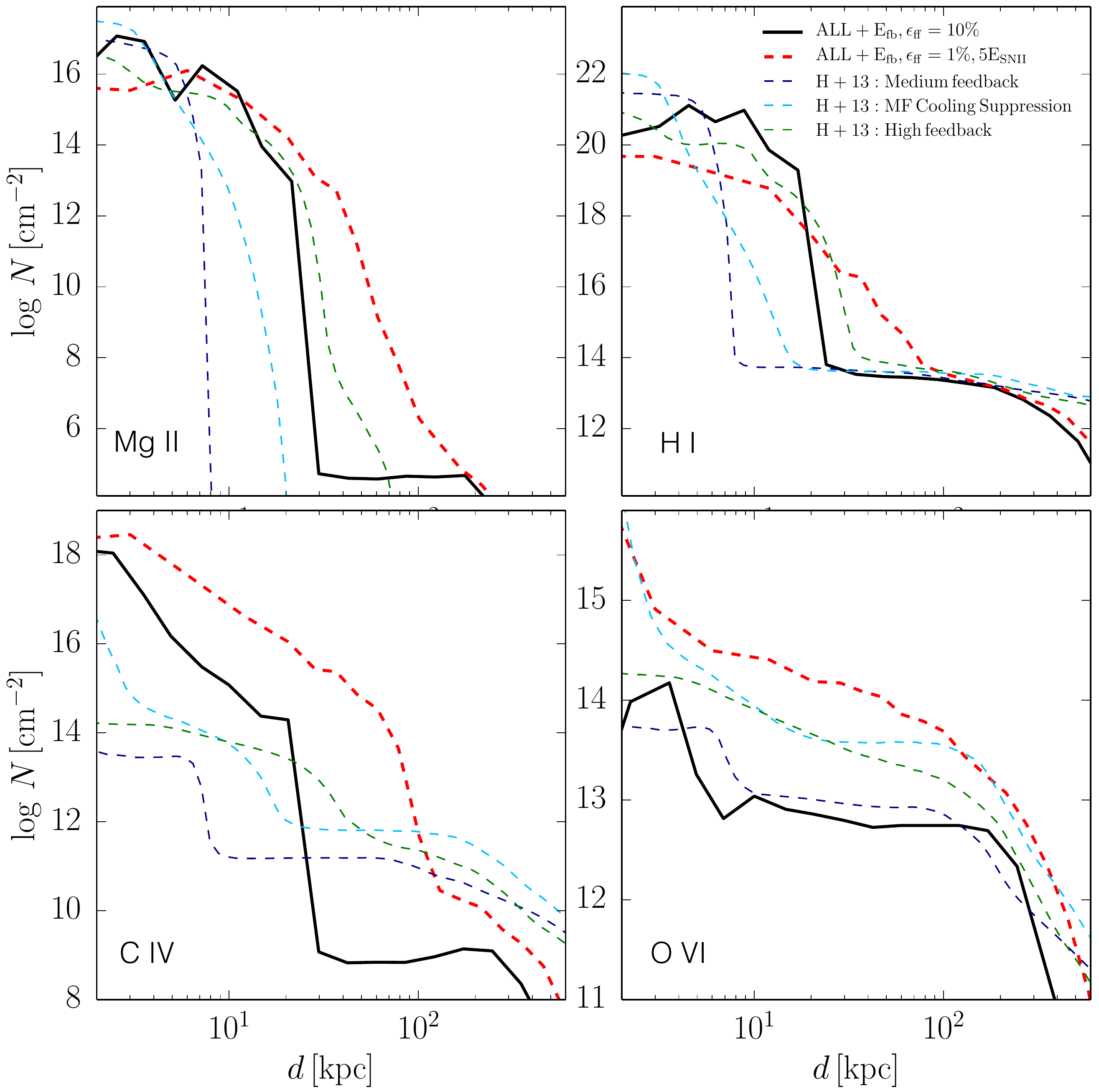}
\caption{Comparisons of predicted column density profiles from low to high ions between our simulations with the profiles from simulations  of \citet[][denoted as H$+$13]{Hummels2013}. The profiles in our fiducial run are closest to the medium feedback run of H+13.}
\label{fig:h13comparison}
\end{center}
\end{figure*}

%-------------------------------------------------------
\subsection{The Effects of Self-Shielding, Non-Equilibrium and Local Starburst Radiation}
%-------------------------------------------------------
In this work, we use three simplifying assumptions in the calculation of ionization abundance: we assume that gas is optically thin, that it is in ionization equilibrium and ionized by the cosmic UVB only (i.e., not accounting for the local radiation from the stars in the simulated galaxy). We qualitatively discuss their effects and implications on the CGM here.

First, self-shielding may increase the column density of low ions compared to our estimate. We find, however, that self-shielding can only change the column density significantly near the star-forming disk where density is high. For example, \cite{rahmati2013} shows that self-shielding affects gas at column densities greater than  $\log N\approx17 $ $\rm cm^{-2}$ and our Figure \ref{fig:proevol} shows that this corresponds to $r\lesssim 25 \rm $ kpc.

Second, the ionizing radiation  from local star-forming regions will likely increase the ratio of low to high ions due to a softer stellar spectrum compared to QSOs. Assuming escape fraction of $f_{\rm esc} = 1-2\%$, the contribution of local radiation is negligible at radii beyond $\approx$ 30-50 kpc \citep[][]{Shen2012}. Note that adding  local radiation would affect our result in the opposite direction from self-shielding because it would increase ionizing radiation in the regions of high gas column density.

Finally, deviations from photo-ionization equilibrium may boost ionization at lower temperatures. However, this effect becomes less important at lower metallicity, especially in the presence of photo-ionization by a radiation background \citep{Oppenheimer2013}, the regime of our analysis.

Given the current small observational statistics, and lack of strong arguments that these effects are significant at large radii, neglecting them should not affect our conclusions significantly. 

%-------------------------------------------------------
\subsection{Comparisons with Previous Studies}
%-------------------------------------------------------
A number of recent studies have explored predictions for the observable properties of the CGM in cosmological simulations of galaxy formation. 
Some studies have focused exclusively on predictions at high redshifts \citep[$z\approx 2-3$][]{Barnes2011,Vandevoort2012,Shen2012,Shen2013,Pallottini2014,Suresh2015,FaucherGiguere2015}. Much of the high-$z$ work has focused on exploring predictions for the distribution and observable properties of neutral hydrogen and dependence of predictions on the implementation of feedback and associated winds  \citep[e.g.,][]{Barnes2011,Vandevoort2012,VandevoortSchaye2012,Pallottini2014,FaucherGiguere2015}. 

Studies that systematically analyzed properties of the CGM at low redshifts used two kinds of simulations. \citet{Ford2013,Ford2014,Ford2015} used a statistical sample of galaxies formed in a simulation of $32h^{-1}\ \rm Mpc$ and $48h^{-1}$ Mpc boxes, but which modelled the wind launching process phenomenologically by imposing a particular specified wind scaling with halo properties (e.g., wind velocity and loading factor) as a function of halo mass.
In these studies the authors have placed particular emphasis on the origin and dynamical cycles of the CGM. 
\citet{Ford2013,Ford2014} showed that low ions, such as Mg\,II, trace dense gas close to galaxies that were part of recent outflows and which will re-accrete onto a disk on a $\sim$Gyr time scale, while high ions, such as O\,VI, have more extended distributions and originate from ancient outflows \citep[see also][]{Shen2012}. 
Our results are in qualitative agreement with the results of these studies. Moreover, the column density profiles presented in \citet{Ford2014} and \citet{Ford2015} have a form in qualitative agreement with results of our simulations. In particular, their profiles are approximately exponential 
in the inner regions, with some ions exhibiting two exponential component profiles \citep[see, e.g., Fig. 10 in][and Fig. 7 in \citeauthor{Ford2014} \citeyear{Ford2014}]{Ford2013}. 

In the second type of simulations, which includes simulations presented in this work, formation of individual galaxies was modelled using different numerical codes ({\tt Enzo}, {\tt RAMSES}, {\tt Gasoline}) in the ``zoom-in'' simulation where all mass and spatial resolution is focused on a single object \citep{Stinson2012,Shen2013,Hummels2013,Roskar2014,Marasco2015}. The higher resolution in the disk region in such simulations allows a more sophisticated treatment of stellar feedback, which allows treating the wind launching process self-consistently without imposing particular 
wind scaling properties. Not surprisingly, the CGM properties in such simulations were shown to be quite sensitive to the details of star formation
and stellar feedback implementation \citep{Hummels2013,Roskar2014,Suresh2015,Marasco2015}. Again, this is in agreement with our results, which indicate strong dependence of the properties of the CGM on the feedback parameters and implementation. We should note that the high sensitivity of the CGM properties to wind modelling is specific to the zoom-in simulations with self-consistent wind launching. \citet{Ford2014,Ford2015} did not find 
significant differences in the CGM produced by the different phenomenological wind models used in their simulations.  

In Figure \ref{fig:h13comparison}, we compare the CGM profiles from two of our simulations with SN feedback with the profiles of \cite{Hummels2013}.\footnote{The profiles are taken from the website of the authors: {\tt http://chummels.org/CGM.html}}  The figure shows that the profiles from our {\tt ALL\_Efb\_e010} are in good agreement with the ``Medium Feedback'' model of \citet{Hummels2013}, which produce a CGM inconsistent with observations. \citet{Hummels2013} found that artificially delaying cooling after gas is heated by SNe makes
wind launching more efficient and produces a more extended CGM halo, although the extent of the ion distribution is still somewhat smaller than
indicated by observations. In all transitions, our model {\tt ALL\_Efb\_e001\_5ESN} can be viewed as a similar variation of the feedback parameters, although instead of suppressing cooling, it assumes larger energy release per supernova. Physically, this can correspond to a larger fraction of thermal energy injected by a supernova retained by the gas or to top-heavy IMF.  Our profiles for this simulation are indeed closest to the ``high feedback'' 
run of \citet{Hummels2013}, although our profiles are somewhat more extended and are thus in better agreement with observations. 

One may question the physical validity of the cooling delay or increasing energy per supernova. However, we think that these choices may approximate some other physical processes that result in more efficient wind driving. For example, when we include cosmic rays with isotropic
treatment of their diffusion, simulations produce CGM profiles quite close to those of the {\tt ALL\_Efb\_e001\_5ESN} run, albeit without assuming 
larger energy per supernova. Overall, these results (and previous studies that produce realistic CGM) show that feedback processes that contribute to wind launching should be quite efficient. 

As we discussed in the previous section, the column density profiles of the CGM ions are predicted to scale linearly with halo scale radius, $r_s$, although we could naively expect that strong feedback could break such self-similar scaling with halo parameters. Other recent studies have reported similar findings. For example, \citet{VandevoortSchaye2012} show that outskirts of gaseous halos of galaxies in the OWLS simulation evolve as expected from the ``virial'' scaling relations. \cite{Pallottini2014} find that CGM profiles around high-redshift galaxies ($z=4$) are self-similar once scaled with $R_{\rm vir}$.
They have found that the HI column density profiles as a function of normalized impact parameter, $d/R_{\rm vir}$, have power law form, while we find exponential form (or double exponential) for all ions and neutral hydrogen absorbers. Our results indicate that column density profiles of the CGM ions do exhibit self-similarity, but they scale better with halo scale radius, $r_s$, rather than with the virial radius. 

%---------------------------
\section{Conclusions}
\label{sec:conclusions}
%---------------------------

We presented a detailed analysis of the observable properties of the circumgalactic medium in a suite of high-resolution cosmological galaxy re-simulations of a Milky-Way sized halo with a variety of star-formation and feedback models. We have also contrasted these predictions with a large set of existing observations of absorbers around galaxies of different mass from $z\approx 0$ to $z\approx 2$.   Our specific findings can be summarized as follows. 

\begin{enumerate}

	\item[\textbf{1.}] Our simulations indicate that the CGM probed by absorbers of different ions arise in gas that was ejected from the disk by stellar feedback. At low redshifts, the outflows in the MW-sized halo modelled in our simulations are not sufficiently energetic to leave the halo and 
	instead produce plumes that reach a finite radius and then turn around to infall back onto the disk. We show that low ionization absorbers originate in such fountain outflows (Fig. \ref{fig:Nmaplow} and Fig. \ref{fig:Nmapinterm}). 

	\item[\textbf{2.}] We find that the column density profiles of various ions and of neutral hydrogen in our simulations are an exponential function of the impact parameter (see Section \ref{sec:exppro}). Ions with higher ionization energy have more extended profiles with the  scale height of the exponential distribution tightly correlated with the ion ionization energy (see Fig. \ref{fig:hEion}, eqs. \ref{eq:mbfitvir}-\ref{eq:mbfitrs}): $h_s \propto E_{\rm ion}^{0.74}$.  At $z \approx 0$, the scale height of warm gas traced by low-ionization species, such as Mg\,II and C\,IV, range in $\approx 0.03-0.07 R_{\rm vir}$, while higher ionization species, such as O\,VI and Ne\, VIII, have scale heights of $\approx 0.32-0.45R_{\rm vir}$ (Table \ref{tab:Eion} and Fig. \ref{fig:hEion}).   
	
	\item[\textbf{3.}] The predicted correlation of scale height with ionization energy of ions is in qualitative and quantitative agreement with observations. In particular, one of our simulations reproduces the radii where observations show a sharp turnover in the column density distribution for different ions  (Fig. \ref{fig:NdRone}, \ref{fig:NdRtwo} and \ref{fig:NdRthree}). We emphasize, however, that this turnover is not due to a sharp boundary, but is a manifestation of the continuous exponential column density profile. (Fig. \ref{fig:Wrevol}).
	
	\item[\textbf{4.}] We find that the total gas column density profile can be approximated by a double exponential profile. The outer exponential distribution is well traced by the distribution of all ions, although column densities of low ions at these radii are predicted to be much below current sensitivity limits. However, the model predicts that the observed column densities of higher ions track the shape of the total gas column density profile at large $r$ quite closely.  

	\item[\textbf{5.}] We find that the physical extent of the CGM in our simulations evolves slower than the virial radius at $z\leq 2$. We show that column density profiles, in the simulations that match the observed CGM, evolve similarly to the halo scale radius, $r_s$. Thus, column density profiles of galaxies at different redshifts can be rescaled using $r_s$ corresponding to their halo mass. This reveals a \textit{self-similarity} in the simulated and observed CGM across four decades of stellar mass \textit{and} 11 billion years in cosmic time (Fig \ref{fig:proevol} and Fig. \ref{fig:Wrevol}).  

	\item[\textbf{6.}] We show that the addition of supernova-produced cosmic ray fluid and associated pressure on the gas produces a CGM with profiles close to low-$z$ observations. The CGM in this case is much more uniform and less patchy compared to the simulations without cosmic rays. These differences can potentially be tested by future observational measurements of the covering fraction profiles.  Overall, we find that CR-driven winds in the galactic halo contain cooler gas ($T < 10^5$ K) compared to winds driven by supernova feedback without cosmic rays in agreement with previous studies (Fig. \ref{fig:MgIIpro}, \ref{fig:allpro} and \ref{fig:coverfrac}). 
\end{enumerate}

All but one simulation presented here reproduce the star formation history expected for a typical $\approx L_*$ galaxy reasonably well. In agreement with other recent studies, we show that properties of the CGM are quite sensitive to the details of the star formation--feedback loop. For example, the fiducial simulation of \citet[][]{AgertzKravtsov2015,AgertzKravtsov2016}, {\tt ALL\_Efb\_e010}, which produces a very realistic central galaxy with the correct stellar mass, size, angular momentum, rotation curve, bulge-to-disk ratio, stellar and gas surface density profiles fails to reproduce existing observations of the CGM. At the same time, variations of star formation efficiency, energy per supernova, or introducing cosmic ray feedback appear to bring the predicted CGM in better
agreement with observations. This illustrates that the properties of galaxies and the properties of their CGM provide strong {\it complementary} constraints on the processes governing galaxy formation. Our results clearly show that future tests of galaxy formation physics should make use of the growing data set of the CGM measurements. Potentially, information on the hot gas component via the highest ions or direct X-ray imaging could also provide valuable constraints on the models of star formation and feedback. 

\section*{Acknowledgments}

CL is grateful for the fruitful discussions with Hsiao-Wen Chen and Sean Johnson. AK would like to thank the Aspen Center for Physics (supported by NSF grant 1066293) for its hospitality and productive atmosphere and organizers of the workshop ``Physics of Accretion and Feedback in the Circum-Galactic Medium'' in June 2015 for opportunities to discuss results presented here with a wide range of colleagues and for the productive atmosphere during completion of this paper. We are particularly grateful to Mark Voit for useful comments on the possible origin of the correlation between scale height and ionization energy.  We also thank the anonymous referee for helpful comments and improving the presentation of the manuscript.  CL and AK were supported by a NASA ATP grant NNH12ZDA001N, NSF grant AST-1412107, and  by the Kavli Institute for Cosmological Physics at the University of Chicago through grant PHY-1125897 and an endowment from the Kavli Foundation and its founder Fred Kavli. CL is partially supported by NASA Headquarters under the NASA Earth and Space Science Fellowship Program - Grant NNX15AR86H. The simulations presented in this paper have been carried using the Midway cluster at the University of Chicago Research Computing Center, which we acknowledge for support.

%%%%%%%%%%%%%%%%%%%%%%%%%%%%%%%%%%%%%%%%%%%%%%%%

\bibliographystyle{mn}
\bibliography{cgm_profiles}

\begin{thebibliography}{101}
\expandafter\ifx\csname natexlab\endcsname\relax\def\natexlab#1{#1}\fi

\bibitem[{{Ackermann} {et~al.}(2013){Ackermann}, {Ajello}, {Allafort}, \& {et
  al.}}]{Ackermann2013}
{Ackermann}, M., {Ajello}, M., {Allafort}, A., \& {et al.} 2013, Science, 339,
  807

\bibitem[{{Agertz} \& {Kravtsov}(2015)}]{AgertzKravtsov2015}
{Agertz}, O., \& {Kravtsov}, A.~V. 2015, \apj, 804, 18

\bibitem[{{Agertz} \& {Kravtsov}(2016)}]{AgertzKravtsov2016}
---. 2016, {\apj} submitted (arxiv/1509.00853)

\bibitem[{{Agertz} {et~al.}(2013){Agertz}, {Kravtsov}, {Leitner}, \&
  {Gnedin}}]{Agertz2013}
{Agertz}, O., {Kravtsov}, A.~V., {Leitner}, S.~N., \& {Gnedin}, N.~Y. 2013,
  \apj, 770, 25

\bibitem[{{Asplund} {et~al.}(2009){Asplund}, {Grevesse}, {Sauval}, \&
  {Scott}}]{Asplund2009}
{Asplund}, M., {Grevesse}, N., {Sauval}, A.~J., \& {Scott}, P. 2009, \araa, 47,
  481

\bibitem[{{Aumer} {et~al.}(2013){Aumer}, {White}, {Naab}, \&
  {Scannapieco}}]{Aumer2013}
{Aumer}, M., {White}, S.~D.~M., {Naab}, T., \& {Scannapieco}, C. 2013, \mnras,
  434, 3142

\bibitem[{{Barai} {et~al.}(2013){Barai}, {Viel}, {Borgani}, {Tescari},
  {Tornatore}, {Dolag}, {Killedar}, {Monaco}, {D'Odorico}, \&
  {Cristiani}}]{Barai2013}
{Barai}, P., {Viel}, M., {Borgani}, S., {et~al.} 2013, \mnras, 430, 3213

\bibitem[{{Barnes} {et~al.}(2011){Barnes}, {Haehnelt}, {Tescari}, \&
  {Viel}}]{Barnes2011}
{Barnes}, L.~A., {Haehnelt}, M.~G., {Tescari}, E., \& {Viel}, M. 2011, \mnras,
  416, 1723

\bibitem[{{Behroozi} {et~al.}(2013){Behroozi}, {Wechsler}, \&
  {Conroy}}]{Behroozi2013}
{Behroozi}, P.~S., {Wechsler}, R.~H., \& {Conroy}, C. 2013, \apj, 770, 57

\bibitem[{{Benson} {et~al.}(2003){Benson}, {Bower}, {Frenk}, {Lacey}, {Baugh},
  \& {Cole}}]{Benson2003}
{Benson}, A.~J., {Bower}, R.~G., {Frenk}, C.~S., {et~al.} 2003, \apj, 599, 38

\bibitem[{{Bernardi} {et~al.}(2013){Bernardi}, {Meert}, {Sheth}, {Vikram},
  {Huertas-Company}, {Mei}, \& {Shankar}}]{Bernardi2013}
{Bernardi}, M., {Meert}, A., {Sheth}, R.~K., {et~al.} 2013, \mnras, 436, 697

\bibitem[{{Booth} {et~al.}(2013){Booth}, {Agertz}, {Kravtsov}, \&
  {Gnedin}}]{Booth2013}
{Booth}, C.~M., {Agertz}, O., {Kravtsov}, A.~V., \& {Gnedin}, N.~Y. 2013,
  \apjl, 777, L16

\bibitem[{{Bordoloi} {et~al.}(2011){Bordoloi}, {Lilly}, {Knobel}, \& {et
  al.}}]{Bordoloi2011}
{Bordoloi}, R., {Lilly}, S.~J., {Knobel}, C., \& {et al.} 2011, \apj, 743, 10

\bibitem[{{Bordoloi} {et~al.}(2014){Bordoloi}, {Tumlinson}, {Werk},
  {Oppenheimer}, {Peeples}, {Prochaska}, {Tripp}, {Katz}, {Dav{\'e}}, {Fox},
  {Thom}, {Ford}, {Weinberg}, {Burchett}, \& {Kollmeier}}]{Bordoloi2014}
{Bordoloi}, R., {Tumlinson}, J., {Werk}, J.~K., {et~al.} 2014, \apj, 796, 136

\bibitem[{{Borthakur} {et~al.}(2013){Borthakur}, {Heckman}, {Strickland},
  {Wild}, \& {Schiminovich}}]{Borthakur2013}
{Borthakur}, S., {Heckman}, T., {Strickland}, D., {Wild}, V., \&
  {Schiminovich}, D. 2013, \apj, 768, 18

\bibitem[{{Borthakur} {et~al.}(2015){Borthakur}, {Heckman}, {Tumlinson},
  {Bordoloi}, {Thom}, {Catinella}, {Schiminovich}, {Dave}, {Kauffmann},
  {Moran}, \& {Saintonge}}]{Borthakur2015}
{Borthakur}, S., {Heckman}, T.~M., {Tumlinson}, J., {et~al.} 2015, {\apj}
  submitted (arXiv:1504.01392)

\bibitem[{{Brook} {et~al.}(2012){Brook}, {Stinson}, {Gibson}, {Wadsley}, \&
  {Quinn}}]{Brook2012}
{Brook}, C.~B., {Stinson}, G., {Gibson}, B.~K., {Wadsley}, J., \& {Quinn}, T.
  2012, \mnras, 424, 1275

\bibitem[{{Bryan} \& {Norman}(1998)}]{BryanNorman1998}
{Bryan}, G.~L., \& {Norman}, M.~L. 1998, \apj, 495, 80

\bibitem[{{Bullock} {et~al.}(2001){Bullock}, {Kolatt}, {Sigad}, {Somerville},
  {Kravtsov}, {Klypin}, {Primack}, \& {Dekel}}]{Bullock2001}
{Bullock}, J.~S., {Kolatt}, T.~S., {Sigad}, Y., {et~al.} 2001, \mnras, 321, 559

\bibitem[{{Caprioli} \& {Spitkovsky}(2014)}]{CaprioliSpitkovsky2014}
{Caprioli}, D., \& {Spitkovsky}, A. 2014, \apj, 783, 91

\bibitem[{{Chen}(2012)}]{Chen2012}
{Chen}, H.-W. 2012, \mnras, 427, 1238

\bibitem[{{Chen} {et~al.}(2010){Chen}, {Helsby}, {Gauthier}, {Shectman},
  {Thompson}, \& {Tinker}}]{Chen2010}
{Chen}, H.-W., {Helsby}, J.~E., {Gauthier}, J.-R., {et~al.} 2010, \apj, 714,
  1521

\bibitem[{{Churchill} {et~al.}(2013){Churchill}, {Nielsen}, {Kacprzak}, \&
  {Trujillo-Gomez}}]{Churchill2013}
{Churchill}, C.~W., {Nielsen}, N.~M., {Kacprzak}, G.~G., \& {Trujillo-Gomez},
  S. 2013, \apjl, 763, L42

\bibitem[{{Correa} {et~al.}(2015){Correa}, {Wyithe}, {Schaye}, \&
  {Duffy}}]{correa:2015}
{Correa}, C.~A., {Wyithe}, J.~S.~B., {Schaye}, J., \& {Duffy}, A.~R. 2015,
  arXiv:1502.00391

\bibitem[{{Crighton} {et~al.}(2015){Crighton}, {Hennawi}, {Simcoe}, {Cooksey},
  {Murphy}, {Fumagalli}, {Prochaska}, \& {Shanks}}]{Crighton2015}
{Crighton}, N.~H.~M., {Hennawi}, J.~F., {Simcoe}, R.~A., {et~al.} 2015, \mnras,
  446, 18

\bibitem[{{Cuesta} {et~al.}(2008){Cuesta}, {Prada}, {Klypin}, \&
  {Moles}}]{Cuesta2008}
{Cuesta}, A.~J., {Prada}, F., {Klypin}, A., \& {Moles}, M. 2008, \mnras, 389,
  385

\bibitem[{{Davis} {et~al.}(2015){Davis}, {Keeney}, {Danforth}, \&
  {Stocke}}]{Davis2015}
{Davis}, J.~D., {Keeney}, B.~A., {Danforth}, C.~W., \& {Stocke}, J.~T. 2015,
  {\apj} submitted (arXiv/1506.04095)

\bibitem[{{Dekel} \& {Silk}(1986)}]{DekelSilk1986}
{Dekel}, A., \& {Silk}, J. 1986, \apj, 303, 39

\bibitem[{{Diemand} {et~al.}(2007){Diemand}, {Kuhlen}, \&
  {Madau}}]{diemand_vl_2007}
{Diemand}, J., {Kuhlen}, M., \& {Madau}, P. 2007, \apj, 667, 859

\bibitem[{{Diemer} \& {Kravtsov}(2014)}]{DiemerKravtsov2014}
{Diemer}, B., \& {Kravtsov}, A.~V. 2014, \apj, 789, 1

\bibitem[{{Diemer} \& {Kravtsov}(2015)}]{DiemerKravtsov2015}
---. 2015, \apj, 799, 108

\bibitem[{{Diemer} {et~al.}(2013){Diemer}, {More}, \& {Kravtsov}}]{Diemer2013}
{Diemer}, B., {More}, S., \& {Kravtsov}, A.~V. 2013, \apj, 766, 25

\bibitem[{{Draine}(2011)}]{Draine2011}
{Draine}, B.~T. 2011, {Physics of the Interstellar and Intergalactic Medium}

\bibitem[{{Dubois} \& {Teyssier}(2008)}]{DuboisTeyssier2008}
{Dubois}, Y., \& {Teyssier}, R. 2008, \aap, 477, 79

\bibitem[{{Efstathiou}(2000)}]{Efstathiou2000}
{Efstathiou}, G. 2000, \mnras, 317, 697

\bibitem[{{Faucher-Gigu{\`e}re} {et~al.}(2015){Faucher-Gigu{\`e}re}, {Hopkins},
  {Kere{\v s}}, {Muratov}, {Quataert}, \& {Murray}}]{FaucherGiguere2015}
{Faucher-Gigu{\`e}re}, C.-A., {Hopkins}, P.~F., {Kere{\v s}}, D., {et~al.}
  2015, \mnras, 449, 987

\bibitem[{{Feldmann} {et~al.}(2011){Feldmann}, {Carollo}, \&
  {Mayer}}]{Feldmann2011}
{Feldmann}, R., {Carollo}, C.~M., \& {Mayer}, L. 2011, \apj, 736, 88

\bibitem[{{Ferland} {et~al.}(2013){Ferland}, {Porter}, {van Hoof}, {Williams},
  {Abel}, {Lykins}, {Shaw}, {Henney}, \& {Stancil}}]{Ferland2013}
{Ferland}, G.~J., {Porter}, R.~L., {van Hoof}, P.~A.~M., {et~al.} 2013, \rmxaa,
  49, 137

\bibitem[{{Ford} {et~al.}(2014){Ford}, {Dav{\'e}}, {Oppenheimer}, {Katz},
  {Kollmeier}, {Thompson}, \& {Weinberg}}]{Ford2014}
{Ford}, A.~B., {Dav{\'e}}, R., {Oppenheimer}, B.~D., {et~al.} 2014, \mnras,
  444, 1260

\bibitem[{{Ford} {et~al.}(2013){Ford}, {Oppenheimer}, {Dav{\'e}}, {Katz},
  {Kollmeier}, \& {Weinberg}}]{Ford2013}
{Ford}, A.~B., {Oppenheimer}, B.~D., {Dav{\'e}}, R., {et~al.} 2013, \mnras,
  432, 89

\bibitem[{{Ford} {et~al.}(2015){Ford}, {Werk}, {Dave}, {Tumlinson}, {Bordoloi},
  {Katz}, {Kollmeier}, {Oppenheimer}, {Peeples}, {Prochaska}, \&
  {Weinberg}}]{Ford2015}
{Ford}, A.~B., {Werk}, J.~K., {Dave}, R., {et~al.} 2015, {\mnras} submitted
  (arXiv:1503.02084)

\bibitem[{{Gauthier} {et~al.}(2010){Gauthier}, {Chen}, \&
  {Tinker}}]{Gauthier2010}
{Gauthier}, J.-R., {Chen}, H.-W., \& {Tinker}, J.~L. 2010, \apj, 716, 1263

\bibitem[{{Green} {et~al.}(2012){Green}, {Froning}, {Osterman}, {Ebbets},
  {Heap}, {Leitherer}, {Linsky}, {Savage}, {Sembach}, {Shull}, {Siegmund},
  {Snow}, {Spencer}, {Stern}, {Stocke}, {Welsh}, {B{\'e}land}, {Burgh},
  {Danforth}, {France}, {Keeney}, {McPhate}, {Penton}, {Andrews},
  {Brownsberger}, {Morse}, \& {Wilkinson}}]{Green2012}
{Green}, J.~C., {Froning}, C.~S., {Osterman}, S., {et~al.} 2012, \apj, 744, 60

\bibitem[{{Haardt} \& {Madau}(2012)}]{HaardtMadau2012}
{Haardt}, F., \& {Madau}, P. 2012, \apj, 746, 125

\bibitem[{{Hopkins} {et~al.}(2014){Hopkins}, {Kere{\v s}}, {O{\~n}orbe},
  {Faucher-Gigu{\`e}re}, {Quataert}, {Murray}, \& {Bullock}}]{Hopkins2014}
{Hopkins}, P.~F., {Kere{\v s}}, D., {O{\~n}orbe}, J., {et~al.} 2014, \mnras,
  445, 581

\bibitem[{{Hopkins} {et~al.}(2011){Hopkins}, {Quataert}, \&
  {Murray}}]{Hopkins2011}
{Hopkins}, P.~F., {Quataert}, E., \& {Murray}, N. 2011, \mnras, 417, 950

\bibitem[{{Hummels} {et~al.}(2013){Hummels}, {Bryan}, {Smith}, \&
  {Turk}}]{Hummels2013}
{Hummels}, C.~B., {Bryan}, G.~L., {Smith}, B.~D., \& {Turk}, M.~J. 2013,
  \mnras, 430, 1548

\bibitem[{{Johnson} {et~al.}(2015){Johnson}, {Chen}, \&
  {Mulchaey}}]{Johnson2015}
{Johnson}, S.~D., {Chen}, H.-W., \& {Mulchaey}, J.~S. 2015, \mnras, 449, 3263

\bibitem[{{Joung} {et~al.}(2012){Joung}, {Bryan}, \& {Putman}}]{Joung2012}
{Joung}, M.~R., {Bryan}, G.~L., \& {Putman}, M.~E. 2012, \apj, 745, 148

\bibitem[{{Komatsu} {et~al.}(2009){Komatsu}, {Dunkley}, {Nolta}, {Bennett},
  {Gold}, {Hinshaw}, {Jarosik}, {Larson}, {Limon}, {Page}, {Spergel},
  {Halpern}, {Hill}, {Kogut}, {Meyer}, {Tucker}, {Weiland}, {Wollack}, \&
  {Wright}}]{Komatsu2009}
{Komatsu}, E., {Dunkley}, J., {Nolta}, M.~R., {et~al.} 2009, \apjs, 180, 330

\bibitem[{{Kravtsov} {et~al.}(2014){Kravtsov}, {Vikhlinin}, \&
  {Meshscheryakov}}]{Kravtsov2014}
{Kravtsov}, A., {Vikhlinin}, A., \& {Meshscheryakov}, A. 2014, arXiv:1401.7329

\bibitem[{{Krumholz} {et~al.}(2009){Krumholz}, {McKee}, \&
  {Tumlinson}}]{Krumholz2009}
{Krumholz}, M.~R., {McKee}, C.~F., \& {Tumlinson}, J. 2009, \apj, 699, 850

\bibitem[{{Leitherer} {et~al.}(1999){Leitherer}, {Schaerer}, {Goldader},
  {Delgado}, {Robert}, {Kune}, {de Mello}, {Devost}, \&
  {Heckman}}]{Leitherer1999}
{Leitherer}, C., {Schaerer}, D., {Goldader}, J.~D., {et~al.} 1999, \apjs, 123,
  3

\bibitem[{{Liang} \& {Chen}(2014)}]{LiangChen2014}
{Liang}, C.~J., \& {Chen}, H.-W. 2014, \mnras, 445, 2061

\bibitem[{Malkov \& Drury(2001)}]{Malkov2001}
Malkov, M., \& Drury, L.~O. 2001, Reports on Progress in Physics, 64, 429

\bibitem[{{Marasco} {et~al.}(2015){Marasco}, {Debattista}, {Fraternali}, {van
  der Hulst}, {Wadsley}, {Quinn}, \& {Ro{\v s}kar}}]{Marasco2015}
{Marasco}, A., {Debattista}, V.~P., {Fraternali}, F., {et~al.} 2015, \mnras,
  451, 4223

\bibitem[{{Mo} {et~al.}(2010){Mo}, {van den Bosch}, \& {White}}]{Mo2010}
{Mo}, H., {van den Bosch}, F.~C., \& {White}, S. 2010, {Galaxy Formation and
  Evolution}

\bibitem[{{More} {et~al.}(2015){More}, {Diemer}, \& {Kravtsov}}]{more2015}
{More}, S., {Diemer}, B., \& {Kravtsov}, A. 2015, arXiv:1504.05591

\bibitem[{{Morton}(2003)}]{Morton2003}
{Morton}, D.~C. 2003, \apjs, 149, 205

\bibitem[{{Muratov} {et~al.}(2015){Muratov}, {Keres}, {Faucher-Giguere},
  {Hopkins}, {Quataert}, \& {Murray}}]{Muratov2015}
{Muratov}, A.~L., {Keres}, D., {Faucher-Giguere}, C.-A., {et~al.} 2015,
  {\mnras} submitted (arxiv:1501.03155)

\bibitem[{{Murray} {et~al.}(2005){Murray}, {Quataert}, \&
  {Thompson}}]{Murray2005}
{Murray}, N., {Quataert}, E., \& {Thompson}, T.~A. 2005, \apj, 618, 569

\bibitem[{{Muzahid}(2014)}]{Muzahid2014}
{Muzahid}, S. 2014, \apj, 784, 5

\bibitem[{{Neto} {et~al.}(2007){Neto}, {Gao}, {Bett}, {Cole}, {Navarro},
  {Frenk}, {White}, {Springel}, \& {Jenkins}}]{Neto2007}
{Neto}, A.~F., {Gao}, L., {Bett}, P., {et~al.} 2007, \mnras, 381, 1450

\bibitem[{{Nielsen} {et~al.}(2013){Nielsen}, {Churchill}, \&
  {Kacprzak}}]{Nielsen2013}
{Nielsen}, N.~M., {Churchill}, C.~W., \& {Kacprzak}, G.~G. 2013, \apj, 776, 115

\bibitem[{{Oppenheimer} {et~al.}(2012){Oppenheimer}, {Dav{\'e}}, {Katz},
  {Kollmeier}, \& {Weinberg}}]{Oppenheimer2012}
{Oppenheimer}, B.~D., {Dav{\'e}}, R., {Katz}, N., {Kollmeier}, J.~A., \&
  {Weinberg}, D.~H. 2012, \mnras, 420, 829

\bibitem[{{Oppenheimer} \& {Schaye}(2013)}]{Oppenheimer2013}
{Oppenheimer}, B.~D., \& {Schaye}, J. 2013, \mnras, 434, 1043

\bibitem[{{Pallottini} {et~al.}(2014){Pallottini}, {Gallerani}, \&
  {Ferrara}}]{Pallottini2014}
{Pallottini}, A., {Gallerani}, S., \& {Ferrara}, A. 2014, \mnras, 444, L105

\bibitem[{{Pitts} \& {Tayler}(1997)}]{PittsTayler1997}
{Pitts}, E., \& {Tayler}, R.~J. 1997, \mnras, 288, 457

\bibitem[{{Prada} {et~al.}(2006){Prada}, {Klypin}, {Simonneau},
  {Betancort-Rijo}, {Patiri}, {Gottl{\"o}ber}, \& {Sanchez-Conde}}]{prada_06}
{Prada}, F., {Klypin}, A.~A., {Simonneau}, E., {et~al.} 2006, \apj, 645, 1001

\bibitem[{{Prochaska} {et~al.}(2011){Prochaska}, {Weiner}, {Chen}, {Mulchaey},
  \& {Cooksey}}]{Prochaska2011}
{Prochaska}, J.~X., {Weiner}, B., {Chen}, H.-W., {Mulchaey}, J., \& {Cooksey},
  K. 2011, \apj, 740, 91

\bibitem[{{Rahmati} {et~al.}(2013){Rahmati}, {Pawlik}, Raicevic, \&
  {Schaye}}]{rahmati2013}
{Rahmati}, A., {Pawlik}, A.~H., Raicevic, M., \& {Schaye}, J. 2013, \mnras,
  430, 2427

\bibitem[{{Rasera} \& {Teyssier}(2006)}]{RaseraTeyssier2006}
{Rasera}, Y., \& {Teyssier}, R. 2006, \aap, 445, 1

\bibitem[{{Reddy} {et~al.}(2012){Reddy}, {Pettini}, {Steidel}, {Shapley},
  {Erb}, \& {Law}}]{Reddy2012}
{Reddy}, N.~A., {Pettini}, M., {Steidel}, C.~C., {et~al.} 2012, \apj, 754, 25

\bibitem[{{Ro{\v s}kar} {et~al.}(2014){Ro{\v s}kar}, {Teyssier}, {Agertz},
  {Wetzstein}, \& {Moore}}]{Roskar2014}
{Ro{\v s}kar}, R., {Teyssier}, R., {Agertz}, O., {Wetzstein}, M., \& {Moore},
  B. 2014, \mnras, 444, 2837

\bibitem[{{Rudie} {et~al.}(2012){Rudie}, {Steidel}, {Trainor}, {Rakic},
  {Bogosavljevi{\'c}}, {Pettini}, {Reddy}, {Shapley}, {Erb}, \&
  {Law}}]{Rudie2012}
{Rudie}, G.~C., {Steidel}, C.~C., {Trainor}, R.~F., {et~al.} 2012, \apj, 750,
  67

\bibitem[{{Rupke} {et~al.}(2005){Rupke}, {Veilleux}, \& {Sanders}}]{Rupke2005}
{Rupke}, D.~S., {Veilleux}, S., \& {Sanders}, D.~B. 2005, \apjs, 160, 115

\bibitem[{{Salem} \& {Bryan}(2014)}]{SalemBryan2014}
{Salem}, M., \& {Bryan}, G.~L. 2014, \mnras, 437, 3312

\bibitem[{{Salem} {et~al.}(2014){Salem}, {Bryan}, \& {Hummels}}]{Salem2014}
{Salem}, M., {Bryan}, G.~L., \& {Hummels}, C. 2014, \apjl, 797, L18

\bibitem[{{Sales} {et~al.}(2012){Sales}, {Navarro}, {Theuns}, {Schaye},
  {White}, {Frenk}, {Crain}, \& {Dalla Vecchia}}]{Sales2012}
{Sales}, L.~V., {Navarro}, J.~F., {Theuns}, T., {et~al.} 2012, \mnras, 423,
  1544

\bibitem[{{Schaye} {et~al.}(2015){Schaye}, {Crain}, {Bower}, {Furlong},
  {Schaller}, {Theuns}, {Dalla Vecchia}, {Frenk}, {McCarthy}, {Helly},
  {Jenkins}, {Rosas-Guevara}, {White}, {Baes}, {Booth}, {Camps}, {Navarro},
  {Qu}, {Rahmati}, {Sawala}, {Thomas}, \& {Trayford}}]{Schaye2015}
{Schaye}, J., {Crain}, R.~A., {Bower}, R.~G., {et~al.} 2015, \mnras, 446, 521

\bibitem[{{Schwartz} \& {Martin}(2004)}]{SchwartzMartin2004}
{Schwartz}, C.~M., \& {Martin}, C.~L. 2004, \apj, 610, 201

\bibitem[{{Schwartz} {et~al.}(2006){Schwartz}, {Martin}, {Chandar},
  {Leitherer}, {Heckman}, \& {Oey}}]{Schwartz2006}
{Schwartz}, C.~M., {Martin}, C.~L., {Chandar}, R., {et~al.} 2006, \apj, 646,
  858

\bibitem[{{Shen} {et~al.}(2012){Shen}, {Madau}, {Aguirre}, {Guedes}, {Mayer},
  \& {Wadsley}}]{Shen2012}
{Shen}, S., {Madau}, P., {Aguirre}, A., {et~al.} 2012, \apj, 760, 50

\bibitem[{{Shen} {et~al.}(2013){Shen}, {Madau}, {Guedes}, {Mayer}, {Prochaska},
  \& {Wadsley}}]{Shen2013}
{Shen}, S., {Madau}, P., {Guedes}, J., {et~al.} 2013, \apj, 765, 89

\bibitem[{{Silk} \& {Rees}(1998)}]{SilkRees1998}
{Silk}, J., \& {Rees}, M.~J. 1998, \aap, 331, L1

\bibitem[{{Smith} {et~al.}(2011){Smith}, {Hallman}, {Shull}, \&
  {O'Shea}}]{Smith2011}
{Smith}, B.~D., {Hallman}, E.~J., {Shull}, J.~M., \& {O'Shea}, B.~W. 2011,
  \apj, 731, 6

\bibitem[{{Socrates} {et~al.}(2008){Socrates}, {Davis}, \&
  {Ramirez-Ruiz}}]{Socrates2008}
{Socrates}, A., {Davis}, S.~W., \& {Ramirez-Ruiz}, E. 2008, \apj, 687, 202

\bibitem[{{Steidel} {et~al.}(2010){Steidel}, {Erb}, {Shapley}, {Pettini},
  {Reddy}, {Bogosavljevi{\'c}}, {Rudie}, \& {Rakic}}]{Steidel2010}
{Steidel}, C.~C., {Erb}, D.~K., {Shapley}, A.~E., {et~al.} 2010, \apj, 717, 289

\bibitem[{{Stinson} {et~al.}(2013){Stinson}, {Brook}, {Macci{\`o}}, {Wadsley},
  {Quinn}, \& {Couchman}}]{Stinson2013}
{Stinson}, G.~S., {Brook}, C., {Macci{\`o}}, A.~V., {et~al.} 2013, \mnras, 428,
  129

\bibitem[{{Stinson} {et~al.}(2012){Stinson}, {Brook}, {Prochaska}, {Hennawi},
  {Shen}, {Wadsley}, {Pontzen}, {Couchman}, {Quinn}, {Macci{\`o}}, \&
  {Gibson}}]{Stinson2012}
{Stinson}, G.~S., {Brook}, C., {Prochaska}, J.~X., {et~al.} 2012, \mnras, 425,
  1270

\bibitem[{{Strong} {et~al.}(2010){Strong}, {Porter}, {Digel},
  {J{\'o}hannesson}, {Martin}, {Moskalenko}, {Murphy}, \&
  {Orlando}}]{Strong2010}
{Strong}, A.~W., {Porter}, T.~A., {Digel}, S.~W., {et~al.} 2010, \apjl, 722,
  L58

\bibitem[{{Suresh} {et~al.}(2015){Suresh}, {Bird}, {Vogelsberger}, {Genel},
  {Torrey}, {Sijacki}, {Springel}, \& {Hernquist}}]{Suresh2015}
{Suresh}, J., {Bird}, S., {Vogelsberger}, M., {et~al.} 2015, \mnras, 448, 895

\bibitem[{{Teyssier}(2002)}]{Teyssier2002}
{Teyssier}, R. 2002, \aap, 385, 337

\bibitem[{{Trujillo-Gomez} {et~al.}(2015){Trujillo-Gomez}, {Klypin},
  {Col{\'{\i}}n}, {Ceverino}, {Arraki}, \& {Primack}}]{TrujilloGomez2015}
{Trujillo-Gomez}, S., {Klypin}, A., {Col{\'{\i}}n}, P., {et~al.} 2015, \mnras,
  446, 1140

\bibitem[{{Tumlinson} {et~al.}(2011){Tumlinson}, {Thom}, {Werk}, {Prochaska},
  {Tripp}, {Weinberg}, {Peeples}, {O'Meara}, {Oppenheimer}, {Meiring}, {Katz},
  {Dav{\'e}}, {Ford}, \& {Sembach}}]{Tumlinson2011}
{Tumlinson}, J., {Thom}, C., {Werk}, J.~K., {et~al.} 2011, Science, 334, 948

\bibitem[{{Turner} {et~al.}(2014){Turner}, {Schaye}, {Steidel}, {Rudie}, \&
  {Strom}}]{Turner2014}
{Turner}, M.~L., {Schaye}, J., {Steidel}, C.~C., {Rudie}, G.~C., \& {Strom},
  A.~L. 2014, \mnras, 445, 794

\bibitem[{{van de Voort} \& {Schaye}(2012)}]{VandevoortSchaye2012}
{van de Voort}, F., \& {Schaye}, J. 2012, \mnras, 423, 2991

\bibitem[{{van de Voort} {et~al.}(2012){van de Voort}, {Schaye}, {Altay}, \&
  {Theuns}}]{Vandevoort2012}
{van de Voort}, F., {Schaye}, J., {Altay}, G., \& {Theuns}, T. 2012, \mnras,
  421, 2809

\bibitem[{{Vogelsberger} {et~al.}(2013){Vogelsberger}, {Genel}, {Sijacki},
  {Torrey}, {Springel}, \& {Hernquist}}]{Vogelsberger2013}
{Vogelsberger}, M., {Genel}, S., {Sijacki}, D., {et~al.} 2013, \mnras, 436,
  3031

\bibitem[{{Werk} {et~al.}(2014){Werk}, {Prochaska}, {Tumlinson}, {Peeples},
  {Tripp}, {Fox}, {Lehner}, {Thom}, {O'Meara}, {Ford}, {Bordoloi}, {Katz},
  {Tejos}, {Oppenheimer}, {Dav{\'e}}, \& {Weinberg}}]{Werk2014}
{Werk}, J.~K., {Prochaska}, J.~X., {Tumlinson}, J., {et~al.} 2014, \apj, 792, 8

\bibitem[{{Zemp}(2014)}]{Zemp2014}
{Zemp}, M. 2014, \apj, 792, 124

\end{thebibliography}
\label{lastpage}

\end{document}